\documentclass[preprint,journal]{vgtc}       

\ifpdf
  \pdfoutput=1\relax                   
  \usepackage{graphicx}                
  \DeclareGraphicsExtensions{.pdf,.png,.jpg,.jpeg} 
\else
  \usepackage{graphicx}                
  \DeclareGraphicsExtensions{.eps}     
\fi%

\graphicspath{{figures/}{pictures/}{images/}{visualisations/}{captures/}{./}} 

\usepackage{microtype}                 
\PassOptionsToPackage{warn}{textcomp}  
\usepackage{textcomp}                  
\usepackage{mathptmx}                  
\usepackage{times}                     
\usepackage{cite}                      
\usepackage{tabu}                      
\usepackage{booktabs}                  

\usepackage[usenames,dvipsnames,svgnames,table]{xcolor}
\usepackage{enumitem}
\usepackage{changepage}
\usepackage{float}
\usepackage{multirow}

\usepackage[final]{changes}

\ieeedoi{10.1109/TVCG.2020.3030450}

\onlineid{0}

\vgtccategory{Research}
\vgtcpapertype{please specify}

\newcommand{\bpstart}[1]{\vspace{0.6mm} \noindent{\textbf{#1.}}}

\newcommand{\system}{FIESTA}
\newcommand{\finding}{\noindent $\bullet$~}

\title{Shared Surfaces and Spaces: Collaborative Data Visualisation in a Co-located Immersive Environment}

\author{Benjamin Lee, Xiaoyun Hu, Maxime Cordeil, Arnaud Prouzeau, Bernhard Jenny and Tim Dwyer}
\authorfooter{
\item
 Benjamin Lee, Xiaoyun Hu, Maxime Cordeil, Arnaud Prouzeau, Bernhard Jenny, and Tim Dwyer are with Monash University. Email: \{benjamin.lee1,annika.hu1,max.cordeil,arnaud.prouzeau,bernie.jenny,\\tim.dwyer\}@monash.edu.
}

\shortauthortitle{Lee \MakeLowercase{\textit{et al.}}: Shared Surfaces and Spaces: Collaborative Data Visualisation in a Co-located Immersive Environment}

\abstract{
Immersive technologies offer new opportunities to support collaborative visual data analysis by providing each collaborator a personal, high-resolution view of a flexible shared visualisation space through a head mounted display.
However, most prior studies of collaborative immersive analytics have focused on how groups interact with surface interfaces such as tabletops and wall displays.
This paper reports on a study in which teams of three co-located participants are given flexible visualisation authoring tools to allow a great deal of control in how they structure their shared workspace. They do so using a prototype system we call \emph{\system{}}: the Free-roaming Immersive Environment to Support Team-based Analysis. Unlike traditional visualisation tools, \system{} allows users to freely position authoring interfaces and visualisation artefacts anywhere in the virtual environment, either on virtual surfaces or suspended within the interaction space.
Our participants solved visual analytics tasks on a multivariate data set, doing so individually and collaboratively by creating a large number of 2D and 3D visualisations.
Their behaviours suggest that the usage of surfaces is coupled with the type of visualisation used, often using walls to organise 2D visualisations, but positioning 3D visualisations in the space around them.
Outside of tightly-coupled collaboration, participants followed social protocols and did not interact with visualisations that did not belong to them even if outside of its owner's personal workspace.
}

\keywords{Immersive analytics, collaboration, virtual reality, qualitative study, multivariate data}

\CCScatlist{ 
 \CCScat{K.6.1}{Management of Computing and Information Systems}%
{Project and People Management}{Life Cycle};
 \CCScat{K.7.m}{The Computing Profession}{Miscellaneous}{Ethics}
}

\teaser{
  \centering
 \includegraphics[width=0.85\linewidth]{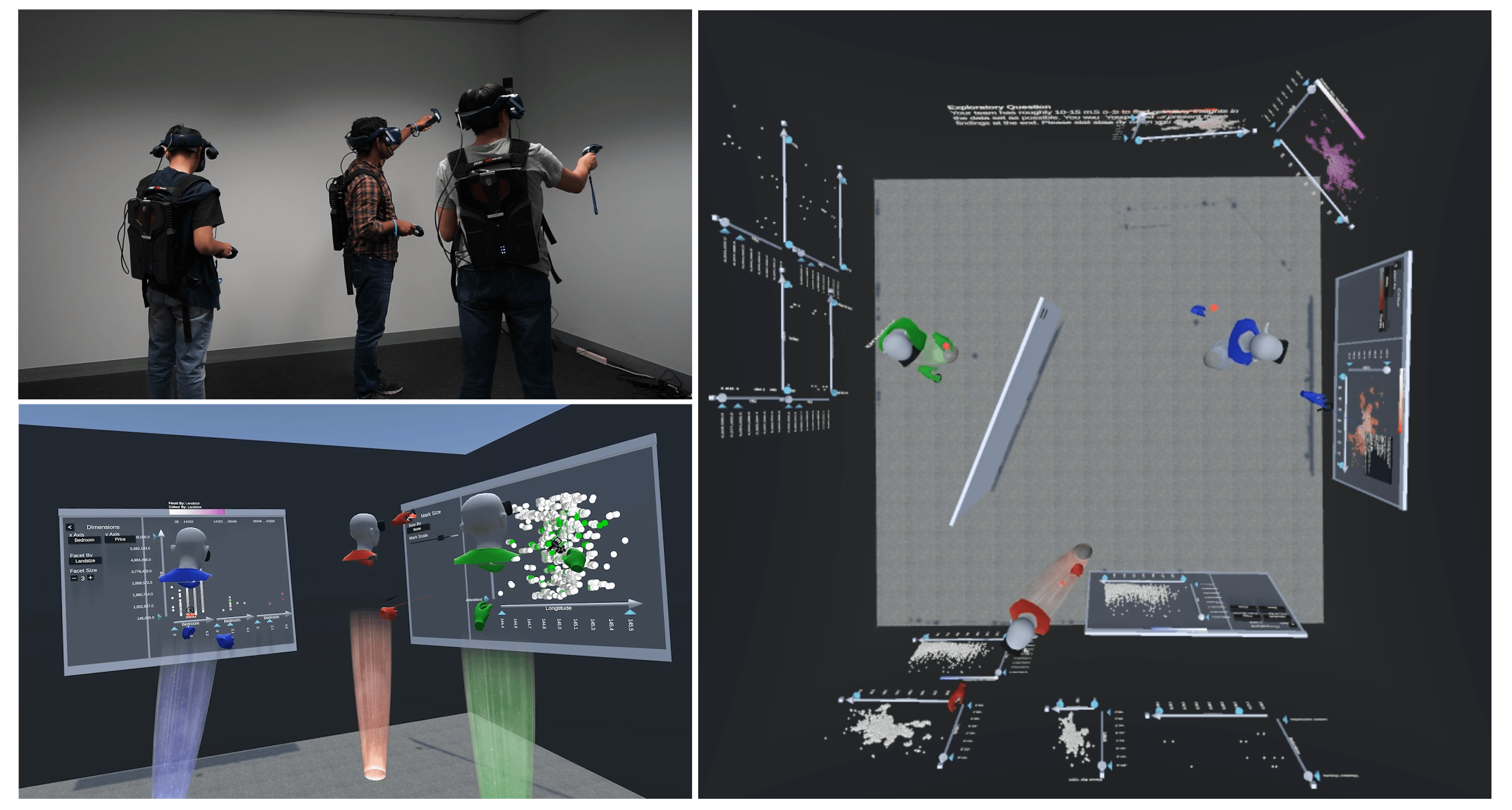}
  \caption{Three participants collaboratively visualising and analysing multidimensional data in the \system{} system, as seen in reality (top left), the virtual environment from the point of view of an observer (bottom left) and a top-down view of the scene from which we analyse their behaviour (right).}
	\label{fig:teaser}
}

\vgtcinsertpkg

\begin{document}


\firstsection{Introduction}

\maketitle

Rapid advances in virtual and augmented reality (VR\replaced{ and }{/}AR) technologies offer exciting possibilities for data visualisation. In the future, the places where we work together---especially for exploring and understanding data---will likely differ from our current desktops and meeting rooms. It is, however, difficult to predict precisely what these places for collaborative immersive analytics will look like.

In lab environments, researchers have invested significant efforts in building collaborative visualisation environments that offer unconventional arrangements of displays, such as wall-sized displays, interactive tabletops, monitors on adjustable stands, and displays projected onto surfaces. More recently, there have been efforts to use AR to extend these displays. An example for individual use is DesignAR, which demonstrates the use of a Microsoft HoloLens with an interactive surface for the creation of 3D models \cite{Reipschlager:2019:DI3}. Specifically for collaborative visualisation, Augmented Reality above the Tabletop (ART) uses immersive parallel coordinates visualisations floating above an interactive tabletop which acts as the interaction surface \cite{Butscher:2018:CTO}. The DataSpace environment \cite{Cavallo:2019:DRH} and its deployed Immersive Insights system \cite{Cavallo:2019:IIH} combines AR with 15 large high-resolution displays mounted on robot arms and a central projection table.
Each of these setups and others like them (as described in Sec.~\ref{sec:related-work}) present interesting examples of specific combinations of display hardware and interaction techniques, but they are a subset of many possible configurations of collaborative environments.
It has been shown \cite{Prouzeau:2017:TOB} that the collaborative behaviour that occurs in such environments is dependent on the arrangement of displays.

In this paper we aim to explore how groups collaborate when given complete freedom to organise their workspace, 
while investigating how users naturally make use of the available surfaces during their analysis.
To do so, we built \system{}: the Free-roaming Immersive Environment to Support Team-based Analysis \cite{Lee:2019:FFR}. \system{} was designed to allow users to create, manipulate and share data visualisations while freely moving in a physically co-located VR environment. It emulates a conventional room with virtual walls and a table, which users may use during their analysis process. Our main contributions are:
\begin{itemize}[leftmargin=*, noitemsep]
    \item Our \system{} prototype, allowing multiple users to create data visualisations in a shared virtual environment -- Sec.~\ref{sec:systemdesign}.
    \item A user study involving the use of \system{} to solve collaborative immersive analytics tasks consisting of 10 groups, each with three participants (30 participants total) -- Sec.~\ref{sec:studyresults} and Sec.~\ref{sec:results}.  To the best of our knowledge, this is the first study observing groups of more than two in a VR data visualisation task.
    \item Our findings from this user study regarding the collaborative use of surfaces and spaces for data visualisation and implications for future development of collaborative immersive analytics -- Sec.~\ref{sec:discussion}.
\end{itemize}
Our observations show that groups were able to perform visual analytics tasks in collaboration. Their use of the surfaces in the environment was influenced by the types of visualisations they created, pinning 2D visualisations on the walls, but placing 3D visualisations egocentrically in the space around them---oftentimes ignoring the table.
In general, we observed that the room was equally divided between participants into territories, with participants respecting ownership of objects during individual work.  
During tightly-coupled collaboration however, participants were willing to use each others' objects regardless of territory.
Finally, participants presented findings to an audience in a variety of ways, but often neglected to account for issues specific to 3D visualisation such as viewing angles and occlusion.
\section{Related work}
\label{sec:related-work}

\subsection{Collaborative Immersive Analytics}
Collaborative immersive analytics has been identified by Billinghurst \emph{et al.} as an emerging field of research at the intersection of InfoVis, CSCW and HCI \cite{Billinghurst:2018:CIA}. 
One of the first immersive platforms for collaborative data visualisation was the CAVE Automatic Virtual Environment (CAVE)~\cite{Cruz-Neira:1993:SPV}. It consisted of a room with screens projected on each wall that immerses users inside the data using stereoscopic technology. Several similar systems have since been developed with improved resolution and tracking systems \cite{DeFanti:2009:STG, Febretti:2013:CHR}. Stereoscopic information visualisation has been shown to improve spatial perception, complex scene understanding, and memory \cite{McIntire:2014:PUS}. Marai \emph{et al.} presented different use cases for the CAVE2, highlighting its benefits for collaborative data analysis \cite{Marai:2016:IIA}. To improve this aspect, Amatriain \emph{et al.}\ presented AlloSphere, a spherical immersive room with a diameter of 10 meters that can accommodate 30 people \cite{Amatriain:2009:AIM}. However, in addition to their high price, these rooms can only provide appropriate stereoscopic vision to a single tracked user.
Cordeil et al.~\cite{Cordeil:2017:ICA} studied immersive collaborative visualisation of 3D graphs between pairs of participants, comparing performance of a CAVE2 with connected VR head-mounted displays (HMDs). They found that performance was highly accurate in both environments, meaning pairs of analysts do not necessarily require a CAVE-style environment for collaborative immersive analytics.



Using VR HMDs, Nguyen \emph{et al.} proposed a preliminary collaborative immersive analytics system in a co-located setting, where pairs of users visualise a star plot in the centre of a virtual room \cite{Nguyen:2019:CDA}. Similar work by the Institute for the Future showcased the Organizational Network in Virtual Reality (ONVR) prototype, which allows groups of remote users to collaboratively interact with a network visualisation in 3D space \cite{IFTF:2018:ONVR}. To the best of our knowledge, these are the only systems involving the use of HMDs for collaborative immersive analytics with no external interfaces or displays. However, both solely focus on specific visualisations, limiting groups to a singular shared visualisation.
Other collaborative systems have sought to integrate AR/VR HMDs with external displays such as tabletops and large displays, oftentimes using these surfaces as shared interaction and work spaces. For example, Butscher \emph{et al.} combined a touch sensitive tabletop and AR to allow users to interact with and manipulate a 3D parallel coordinates visualisation \cite{Butscher:2018:CTO}. Cavallo \emph{et al.} proposed DataSpace, a collaborative room-scale immersive environment composed of AR HMDs and many configurable large displays arranged in a CAVE-like fashion \cite{Cavallo:2019:DRH}. They subsequently presented Immersive Insights, a hybrid analytics system built on the DataSpace environment which integrates touch sensitive displays along the edges of the environment, a touch-enabled projection table in the centre of the room, and an AR data visualisation floating above it for collaborative data analysis \cite{Cavallo:2019:IIH}.
In our work, we aim to understand how larger groups perform collaborative immersive analytics tasks in an unconstrained immersive environment, whereby users can move freely and are not restricted to tabletops or large displays.



%

\subsection{Territoriality in collaborative contexts}
When collaborating in shared environments, people oftentimes divide the workspace into separate territories. Scott \emph{et al.} observed participants working together on a physical tabletop and identified three types of territories which were implicitly created: \textit{personal}, \textit{shared}, and \textit{storage} territories \cite{Scott:2004:TCT}. These territories are dynamic, changing depending on the needs of the activity. When individuals are required to move, Tang \emph{et al.} found that these territories constantly shift in response to their physical positions \cite{Tang:2006:CCO}, as people tend to establish a personal space in the area directly in front of them \cite{Tang:1991:FOS, Kruger:2004:ROT}.
In contrast to tabletops, high-resolution wall displays offer a much larger space for collaborators to share. Jakobsen and Hornb{\ae}k observed that territories were transient, with groups fully sharing the screen without any explicit negotiation \cite{Jakobsen:2014:UCP}. However, instances where participants interacted with an area on the wall display that was directly in front of someone else only occurred during tightly-coupled work.
On the other hand, Liu \emph{et al.} found that pairs of participants work at different areas of the wall display, but only when given facilities to communicate and assist each other across distances \cite{Liu:2016:SIW}.
While territoriality has been studied for tabletops and wall displays, how users move and manage these territories is constrained by the surface. In this work, we investigate a fully unconstrained environment where users can work in the open space around them, and freely utilise any available nearby surfaces.

\subsection{View and window management}
AR\replaced{ and }{/}VR researchers have explored the manner in which views, in our case free-floating data visualisations, can be arranged and managed to evoke spatial memory and cognition.
Robertson \emph{et al.} used an art gallery metaphor to place windows in a 3D space, with the main task window on a `stage' and others along the walls of the gallery \cite{Robertson:2000:TGW}.
An earlier work by Feiner \emph{et al.} showcased an AR system for placing virtual windows in the 3D world around the user \cite{Feiner:1993:WWW}. They defined three types of windows: \textit{surround-fixed windows} positioned along a sphere surrounding the user, \textit{display-fixed windows} positioned directly on the heads up display, and \textit{world-fixed windows} fixed to movable objects in the 3D world.
Ens \emph{et al.} used this concept of body- and world-fixed windows in The Personal Cockpit, with AR windows automatically arranged in a curved body-fixed or flat world-fixed layout \cite{Ens:2014:PCS}. Extending this, Ens \emph{et al.} demonstrated how body-fixed layouts can automatically transition into world-fixed layouts while preserving relative spatial consistency in diverse environments \cite{Ens:2016:SFS}.
While much work has explored how view management can be automated \cite{Bell:2001:VMV, Kruijff:2010:PIA}, there is no agreed upon windowing metaphor for the arrangement of these views \cite{Marriott:2018:IAT}, and little work exploring how users naturally organise these views when performing data visualisation tasks in immersive environments. Most relevant to our work is that by Cordeil \emph{et al.} on ImAxes~\cite{Cordeil:2017:IIA} and the subsequent evaluation by Batch \emph{et al.}~\cite{Batch:2020:TNS}. They found that users would arrange their views egocentrically in a body-fixed like fashion while working \cite{Feiner:1993:WWW, Ens:2014:PCS}, switching to a linear or semi-circular layout for presentation. However, it is uncertain if the introduction of multiple users would influence this behaviour.
\section{A prototype free-roaming collaborative immersive analytics system}
\label{sec:systemdesign}
We designed and developed \system{} to study how groups collaboratively explore data in a free-roaming immersive analytics environment\added{ \cite{Lee:2019:FFR}}. The source code is made publicly available on a GitHub repository\footnote{https://github.com/benjaminchlee/FIESTA}.
We developed \system{} with three fundamental design requirements in mind, which we describe in turn in this section.



\begin{figure}[!t]
    \centering
    \includegraphics[trim = 120 30 190 10, clip, height=4.175cm]{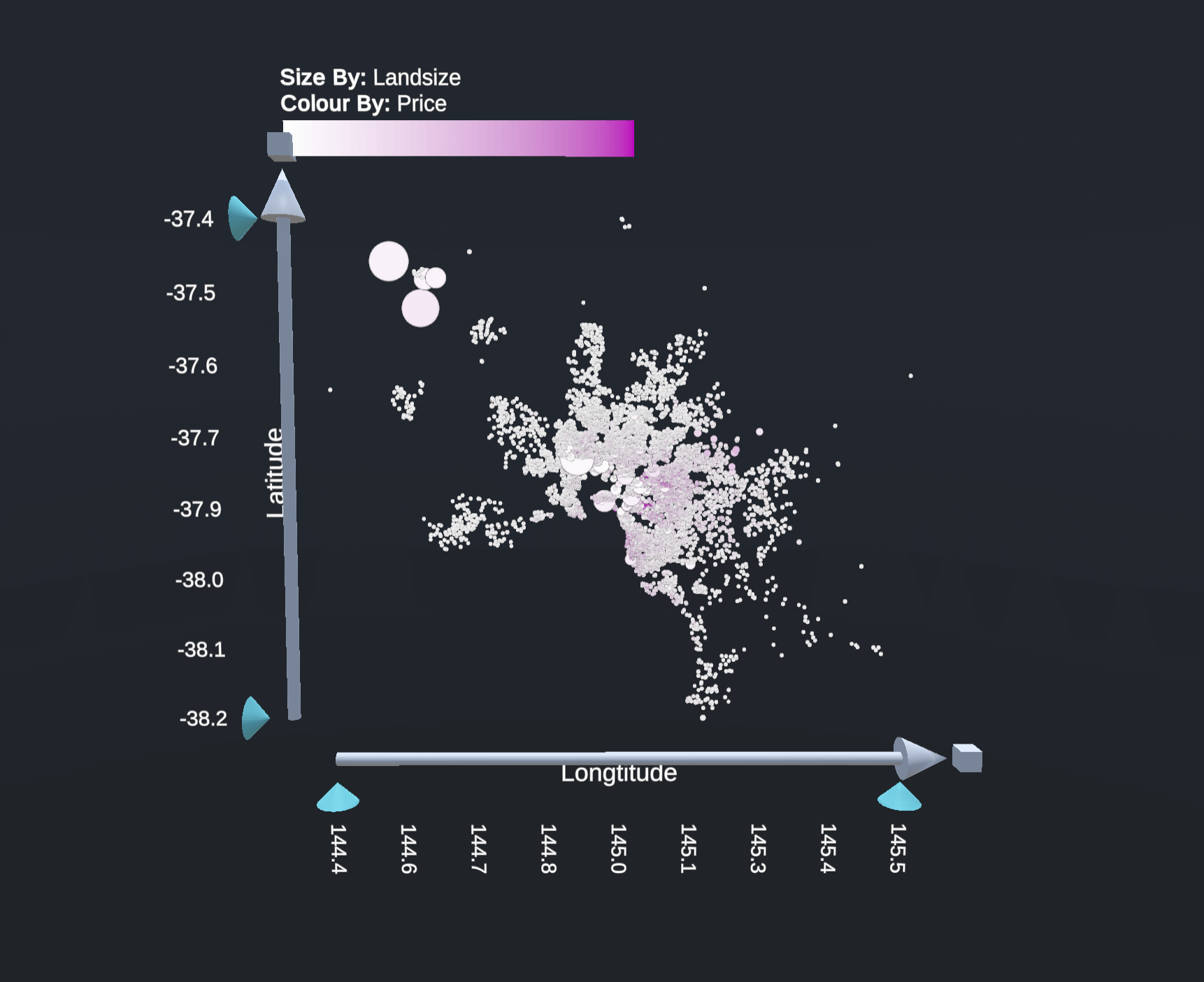}~
    \hspace{-0.15cm}
    \includegraphics[trim = 0 0 0 0, clip, height=4.175cm]{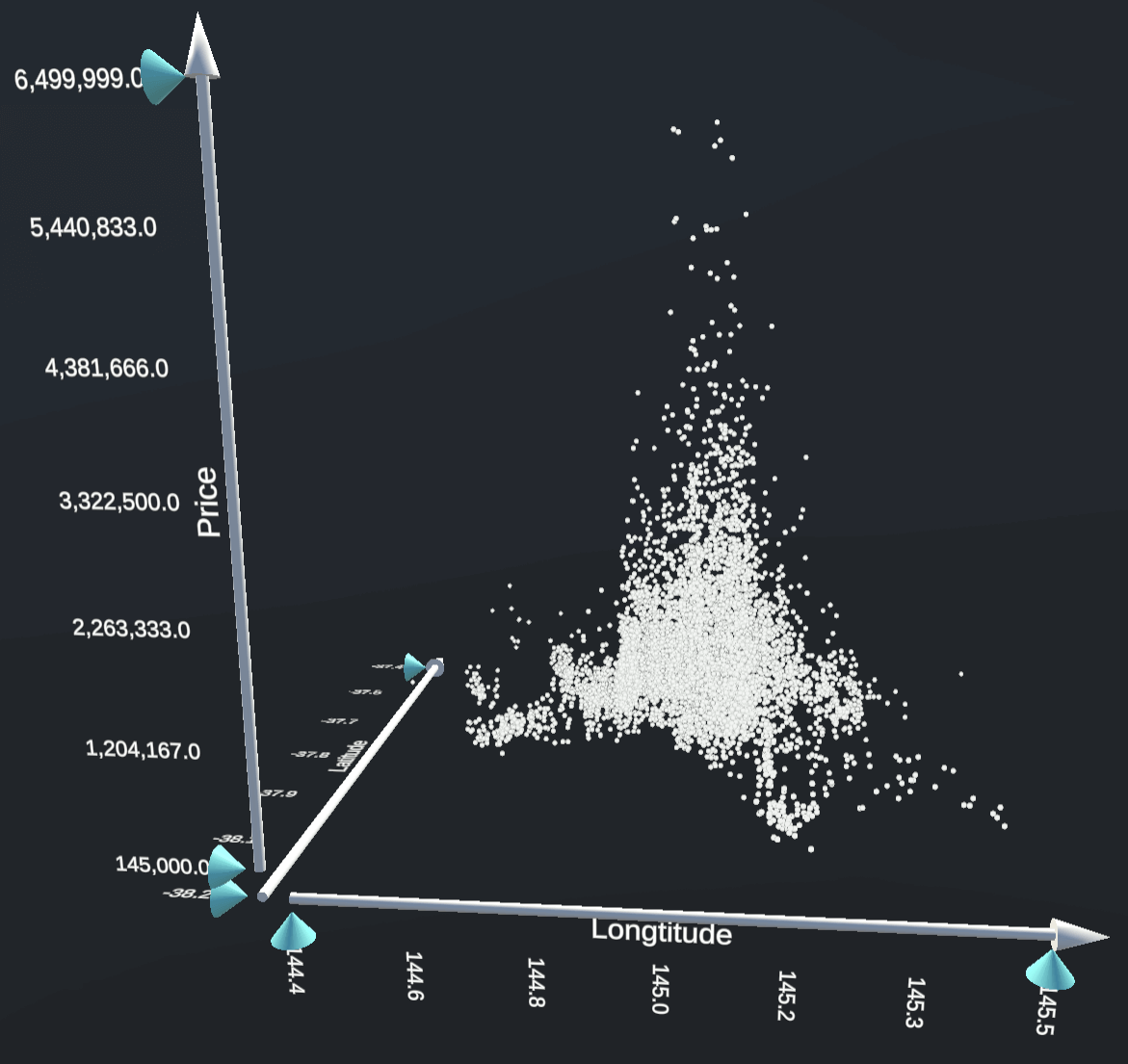} \\
    \vspace{0.05cm}
    \includegraphics[trim = 120 20 180 10, clip, height=3.95cm]{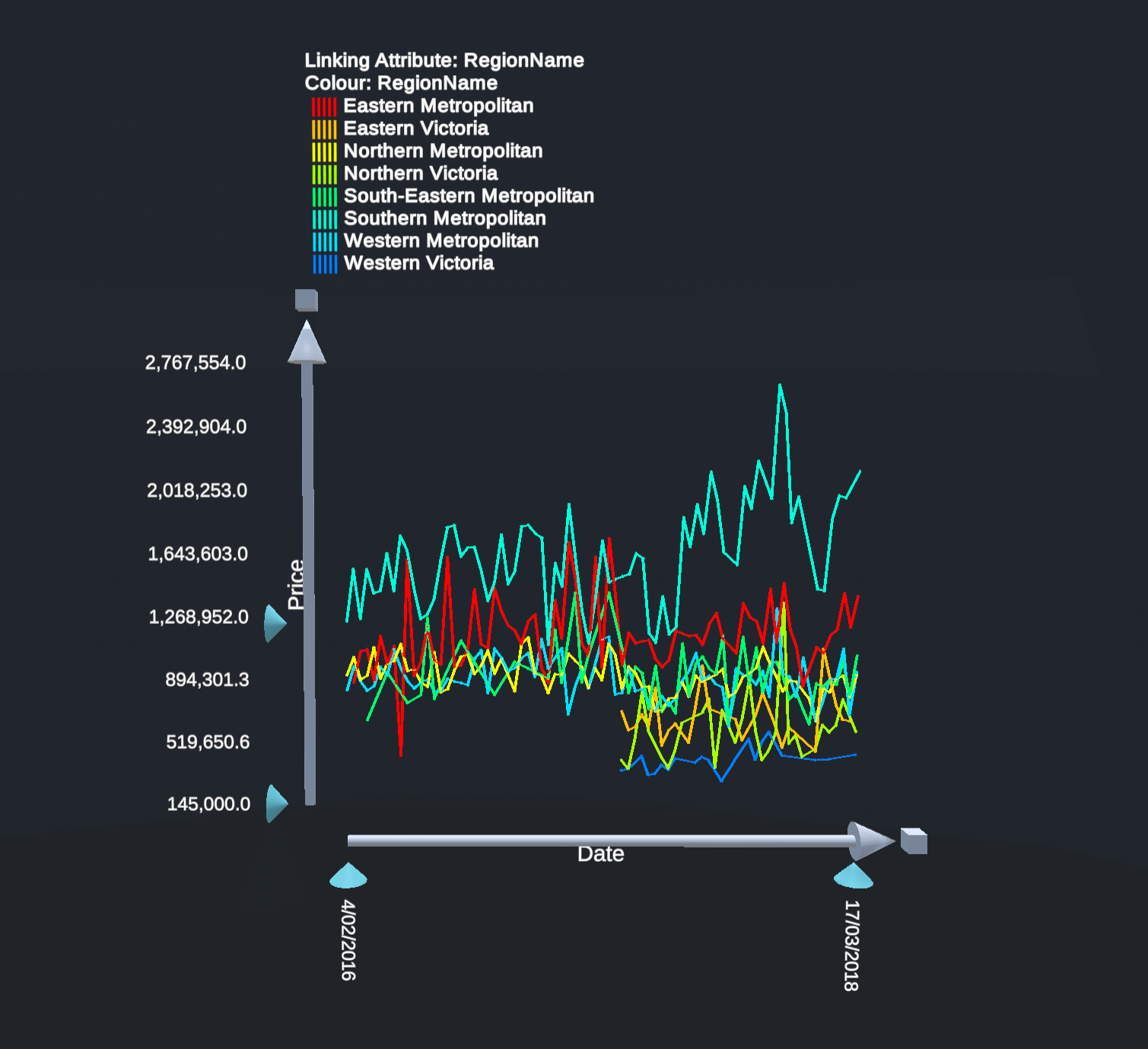} ~
    \hspace{-0.225cm}
    \includegraphics[trim = 100 70 130 30, clip, height=3.95cm]{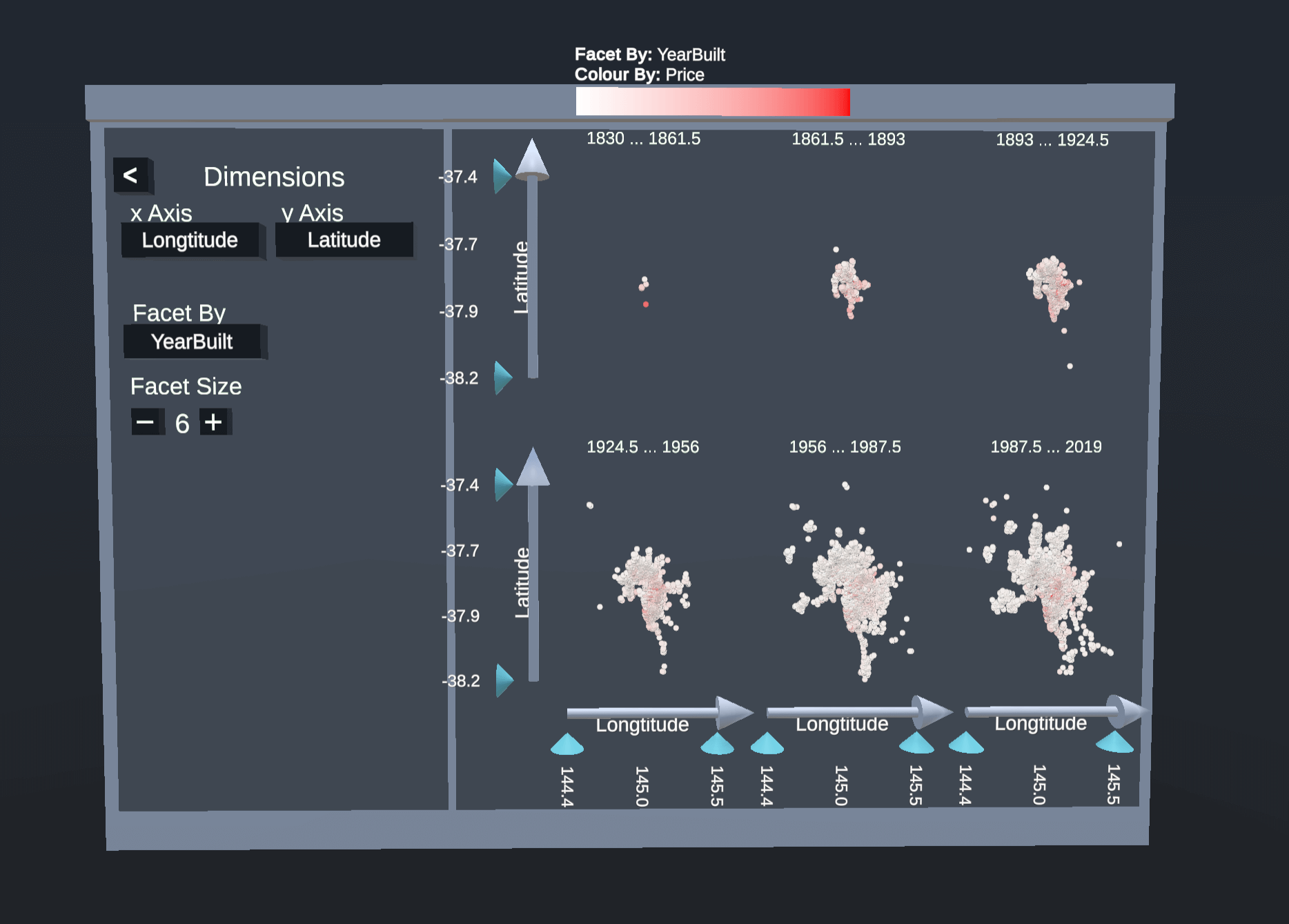}
    \caption{Visualisation styles available with \system{}: 2D scatterplot (top left), 3D scatterplot (top right), time series  (bottom left), faceted 2D scatterplots on the panel (bottom right).}
    \label{fig:visualisation-styles}
\end{figure}

\subsection{Requirement 1: Baseline Data Visualisation System}
Given the novelty of collaborative immersive analytics, we implement a baseline level of functionality expected of any data visualisation system \cite{Heer:2012:Interactive}. We describe these in terms of the key characteristics and functions of \system{}.

\bpstart{Visualisations}
\system{} was built around a customised version of the Immersive Analytics Toolkit (IATK) \cite{Cordeil:2019:IAT}. We use IATK because it is optimised for rendering hundreds of thousands of data points at high frame rates, which is important when multiple users can produce dozens of visualisations each. Users can assemble three main types of 2D and 3D visualisations: scatterplots, faceted scatterplots, and time series (Fig.~\ref{fig:visualisation-styles}). For scatterplots, each data record is represented by a circular billboard. For time series, lines are drawn between billboarded points. Visualisations are static objects that float in 3D space.

\bpstart{Interaction}
We choose user interface elements that are standard across desktop and VR environments to minimise learning time. We adapt 3D UI interaction metaphors \cite{Laviola:2017:3UI} for data visualisation authoring and manipulation: grasping techniques involving direct contact with UI elements at close range, and ranged pointing techniques involving distant interaction with UI elements using a laser pointer.

\bpstart{Authoring}
To create visualisations, users interact with flat panels reminiscent of physical touch panel displays (Fig.~\ref{fig:visualisation-styles}, bottom right). A panel provides a 1.6~m~$\times$~1.1~m 
\replaced{interface}{work space} that can be moved by grasping and releasing \added{it} at the desired location, where it remains suspended in space. One panel is provided for each user. 
The panel is split between a user interface and an accompanying visualisation. The user interface consists of navigable pages which expose the following functions: binding data dimensions to glyph size, colour, $x$, $y$ and (optionally) $z$ position; defining data attributes to build time series; faceting by dimension and adjusting number of facets; and defining colour channels with gradients for continuous attributes and discrete palettes for categorical attributes. Visualisations update in real time when their properties are changed. 2D visualisations are positioned parallel to the panel, with 3D visualisations protruding outwards from it.
All operations utilise conventional user interface elements, including buttons, dropdown menus, sliders, and HSV colour pickers. These are activated by performing a direct grasping or ranged laser pointer action on the interface element.

\begin{figure}[!t]
    \centering
    \includegraphics[height=2.55cm]{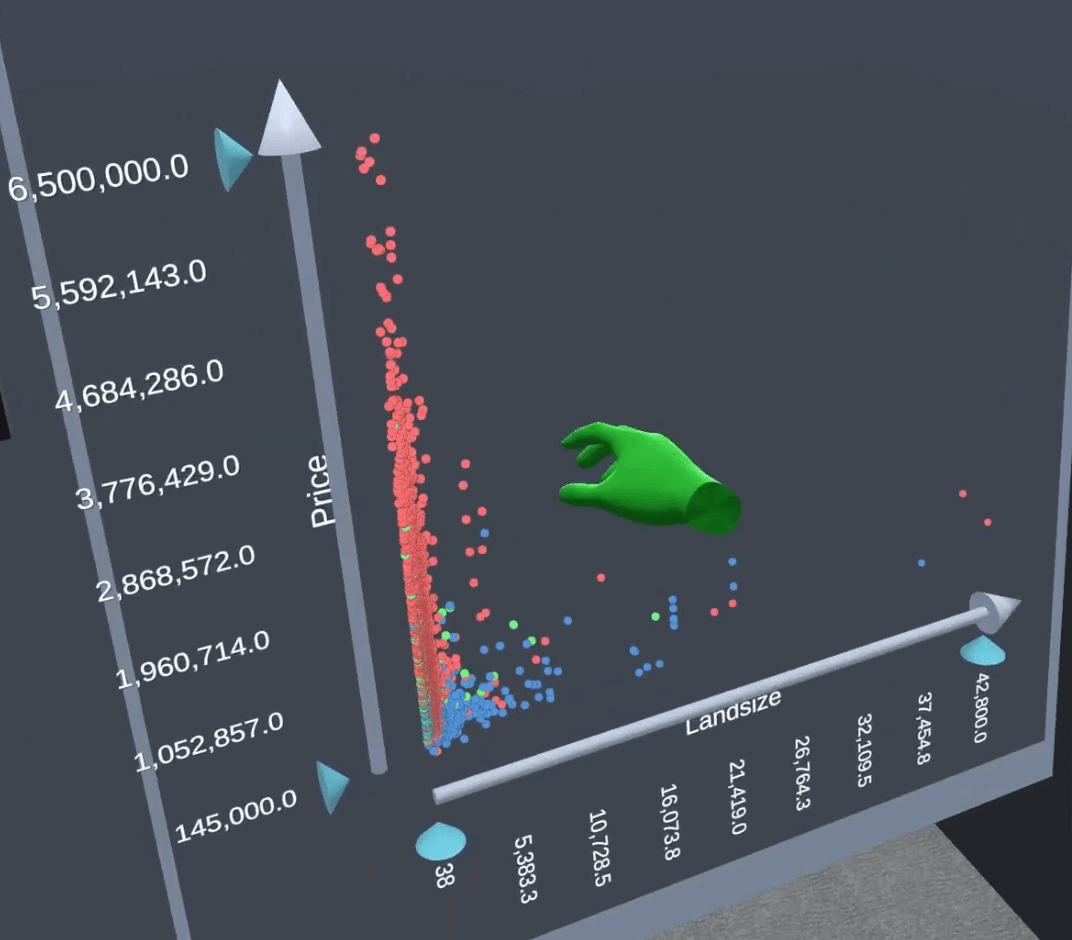}~
    \includegraphics[height=2.55cm]{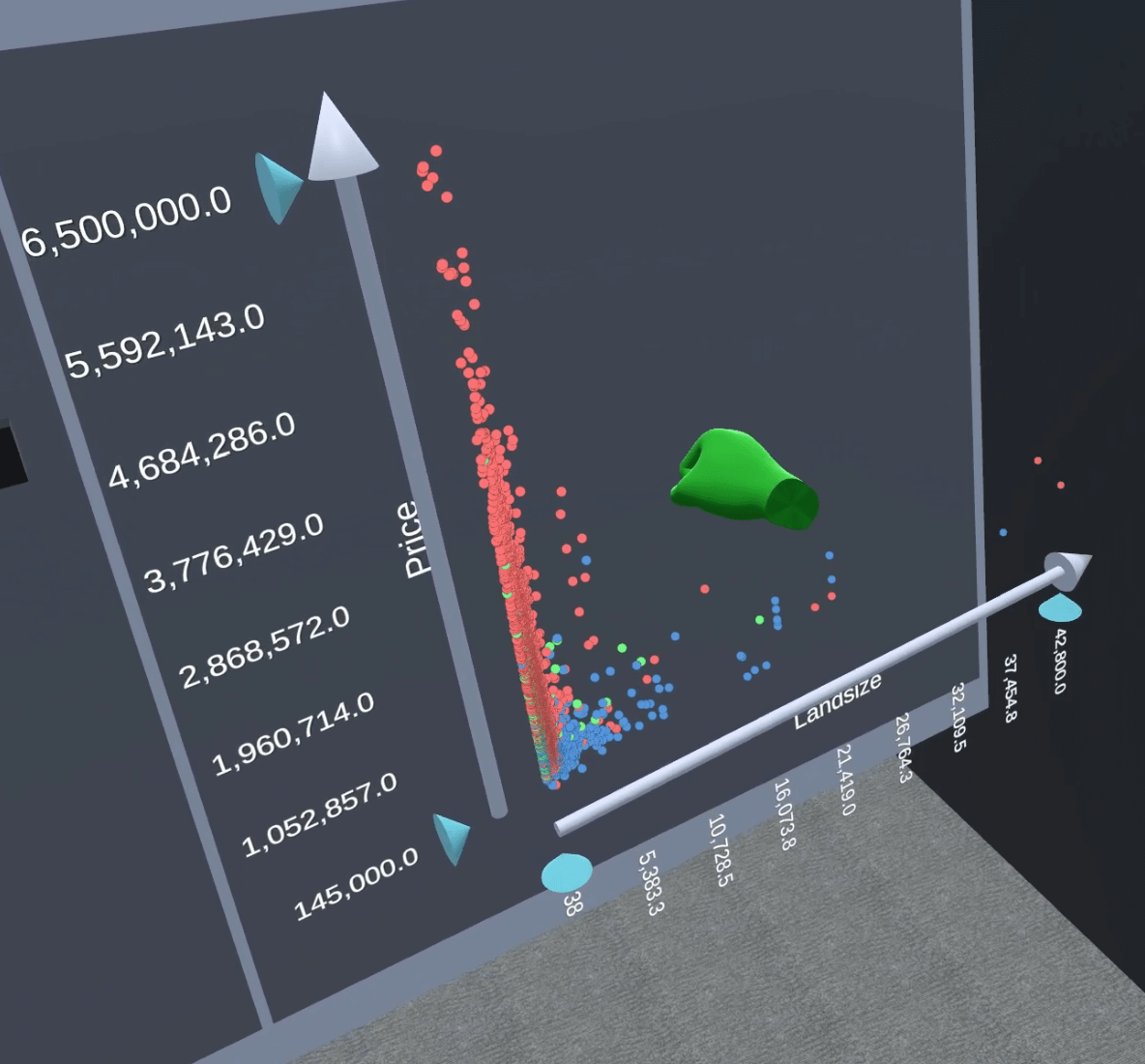}~
    \includegraphics[height=2.55cm]{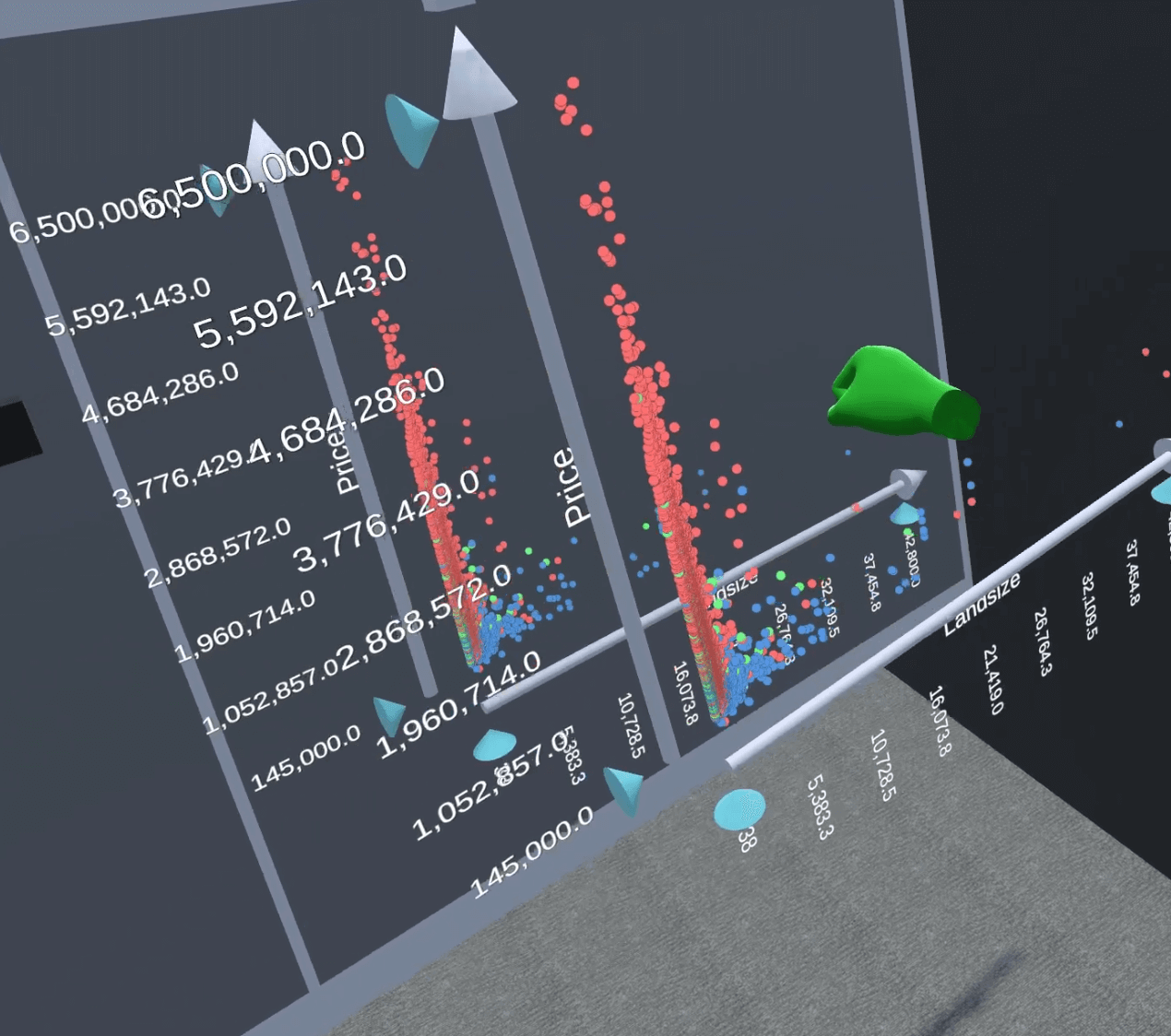}~
    \caption{\system{} uses a tear-out metaphor to duplicate visualisations from panels, allowing them to be freely positioned either in space or on the surfaces of the virtual room.}
    \label{fig:tearout}
\end{figure}

\bpstart{Tearing out visualisations}
After a visualisation is edited on a panel, it can be cloned using a `tearing out' action, grasping it and pulling it away from the panel (Fig.~\ref{fig:tearout}). The copy can then be freely positioned in 3D space through typical grasping actions, with the original snapping back into place on the panel.
Torn out visualisations can be adjusted through grasping interactions on widgets along their axes: resizing the visualisation along a dimension; and rescaling minimum and/or maximum axis ranges to adjust the visible domain of the dimension. Visualisations can be dropped onto a panel for further editing, or destroyed with a `throwing' action aimed at the ground.

\begin{figure}[b]
    \centering
    \includegraphics[height=3.1cm]{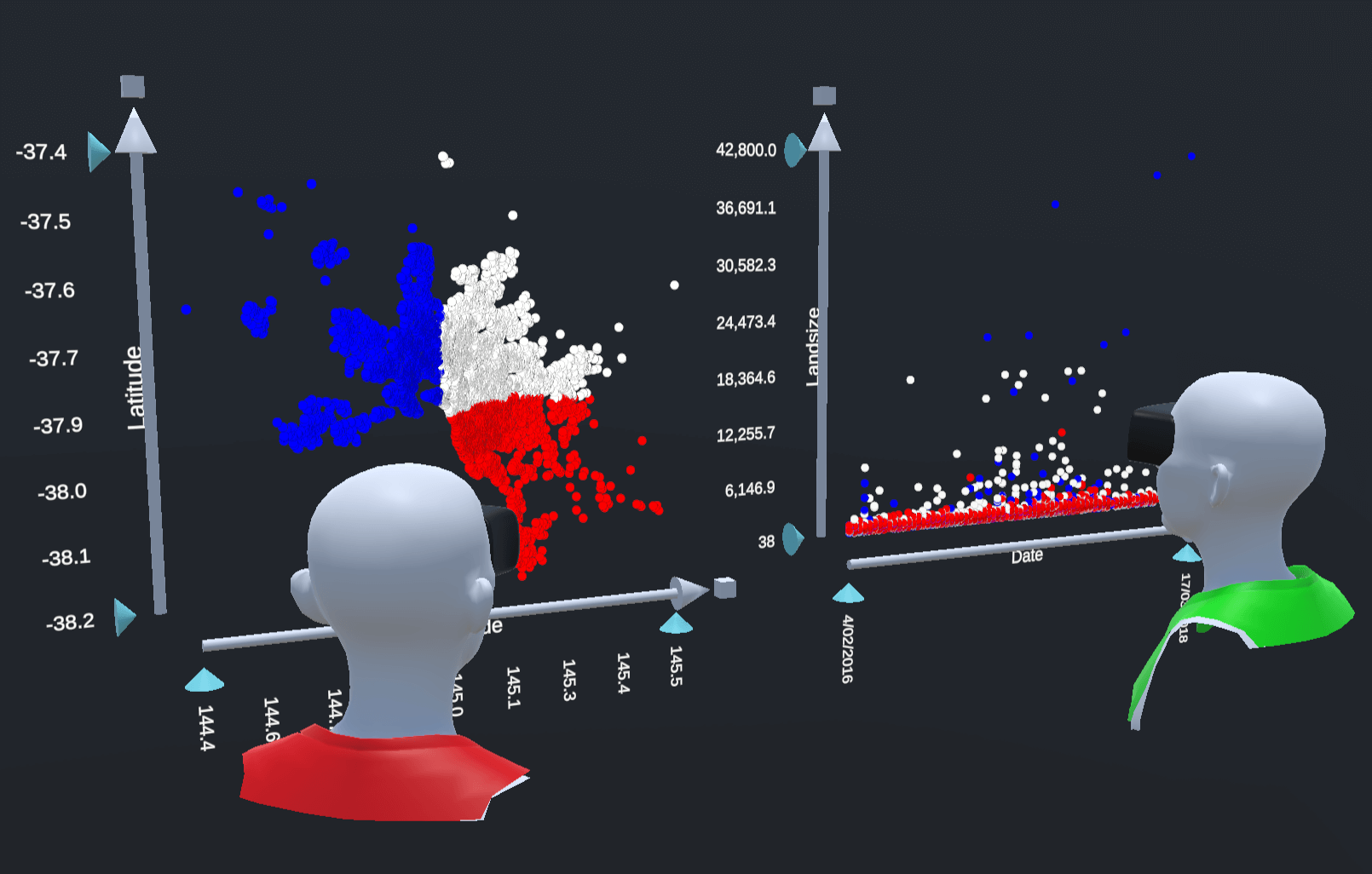}~
    \includegraphics[height=3.1cm]{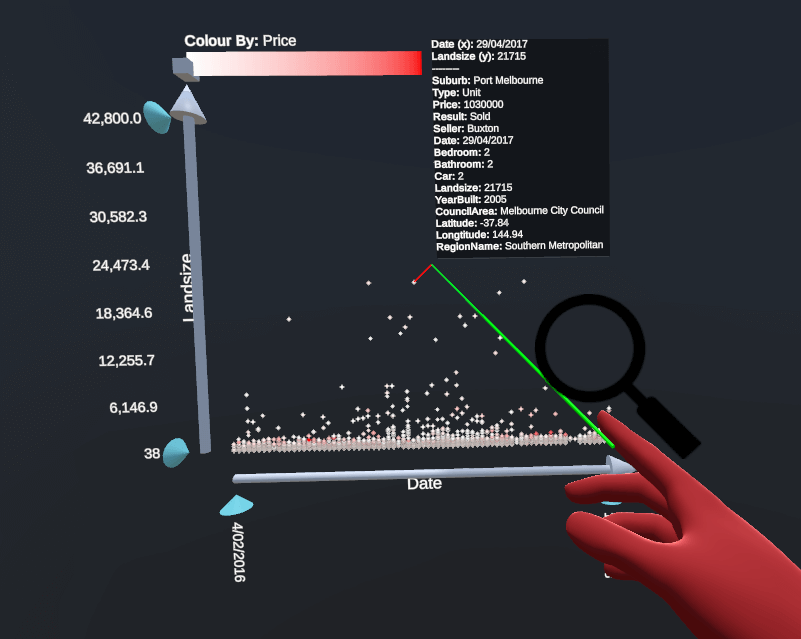}
    \caption{\system{} supports additional data visualisation functionalities: private and shared brushing modes which are linked across all visualisations (left); details on demand to easily inspect data records (right).}
    \label{fig:brushing}
\end{figure}

\bpstart{Pointer, brushing, and annotation tools}
A set of additional tools are available via a spin menu \cite{Gerber:2005:SMM} above the user's offhand which the user can enable: a private brush to make selections that are only visible to the user, a shared brush to make selections which are shown to all users, and a ranged grab which can be used to pull distant visualisations towards the user.
Brushing and selection of data records is linked across all visualisations (Fig.~\ref{fig:brushing}, left).
A pointer appears when a tool is enabled, and is activated whenever the controller's trigger is pulled.
In addition, a details-on-demand tool can be enabled by holding down the touchpad on the user's dominant hand. This shows details of the nearest data record to the pointer (Fig.~\ref{fig:brushing}, right) in an infobox connected by a leader line to the glyph. 
The pointer can also be used to interact with a panel's interface, either using point and click (for buttons) or point and drag (for sliders and pickers).
Lastly, we include basic annotation tools, accessible through floating marker and eraser objects which users can pick up and use. These annotations `attach' to visualisations when drawn on top of their bounding boxes.

\subsection{Requirement 2: Free Roaming Shared Environment}
Our goal was to emulate a physically co-located collaborative environment using VR. We use a tetherless VR setup to prevent tripping and enable free roaming around the environment. Users see each other as virtual avatars aligned to their real-world positions. We use the Oculus Avatar SDK which provides embodied head, torso, and hand models, supporting presence, deixis (i.e.\ pointing, finger gestures), and gaze direction \cite{Heer:2008:DCF}. Each user is uniquely identified by a floating nameplate and avatar colour. The same colour is also used for shared brush selections. This allows users to see the actions of others to support collaborative tasks and information sharing, as well as to avoid physical collisions. \replaced{In addition, we support keyboard and mouse controlled virtual avatars that are visible to VR users, which can be used as a remote audience member (described further in Sec.~\ref{sec:studyresults}).}{In addition, we include an optional virtual avatar (referred to as the audience) which is visible to VR users and controlled by keyboard and mouse, described further in Sec.~\ref{ssc:study-design}.}

Sun \emph{et al.} suggest private workspaces increase the need for communication in order for collaborators to synchronise as a group \cite{Sun:2019:CVA}. To mitigate this and to promote workspace awareness \cite{Gutwin:2002:DFW}, we make everything visible to all users at all times, including pointers, details-on-demand tooltips, etc.
The exception is the private brush tool, as shared brushing can disturb other users \cite{Prouzeau:2017:EMU}.
While each user is given their own panel at the start of the session, \system{} allows users to interact with any object in the environment, \replaced{supporting simultaneous interaction of panels and visualisations with no explicit ownership system}{supporting both individual and shared workflows on the same panel}. 

\subsection{Requirement 3: Room and Surface Affordances}
To explore how users naturally utilise surface affordances in collaborative immersive analytics, we establish a room metaphor similar to prior work \cite{Cruz-Neira:1993:SPV, Cavallo:2019:DRH, Nguyen:2019:CDA} using a standard four-wall room layout. These walls imitate basic interaction with wall-sized displays, allowing users to freely `pin' visualisations on the \replaced{vertical}{flat} surface. Likewise, we incorporate a tabletop metaphor through a square table in the middle of the room (1~m$\times$1~m, 1~m tall). The table allows for visualisations to `rest' on top of it\replaced{ similar to}{, a notion commonly found in} prior work \cite{Butscher:2018:CTO, Cavallo:2019:DRH, Kraus:2020:IIC}. These pinning functions are performed by releasing a visualisation close to the surface, or by physically throwing a visualisation at it. Virtual surfaces are not aligned with any physical objects\deleted{,} meaning users can simply walk through them, but the walls act as a \deleted{preliminary} boundary to prevent collisions with the actual wall\added{s} in the room. 
Because visualisations can be suspended in space, the use of surfaces is inherently optional. However, since past work places great emphasis on interaction with surfaces (e.g.~\cite{Kunert:2019:MWI}) we felt it was important to provide them and to study their effect.


\subsection{Technical Implementation and Deployment}
\label{ssc:technical-implementation}
\system{} was developed using Unity3D. While we have confirmed it to work with remote collaborators via the internet, we chose to deploy the system on three VR `backpacks' with HTC Vive Pros in a 16~m\textsuperscript{2} space (4~m~$\times$~4~m). The VR backpacks are Intel Core i7-7820HK (2.9~GHz, 4 cores) PCs with a Nvidia GeForce GTX 1070 (8~GB) GPU, and 16~GB of RAM \added{VR One PCs from MSI \cite{MSI-VRBackpack}}. We bundle and secure any cables to enable truly free-roaming VR with no tethers or tripping hazards. Four SteamVR 2.0 base-stations were positioned at each corner of the room to minimise tracking issues. The backpack PCs were managed remotely by a fourth desktop PC running an Intel Core i7-7800X CPU (3.5~GHz, 6 cores), Nvidia GeForce GTX 1080 (8~GB) GPU, and 32~GB of RAM. This desktop was positioned outside of the VR space. Networking was done using the Photon Unity Networking engine. All computers were connected to a server running on the desktop PC on a dedicated WLAN to minimise external interference and latency. A server update rate of 30~Hz was used to balance between smooth synchronisation of objects and battery life of the backpacks.
\section{User Study}
\label{sec:studyresults}
The purpose of this work is to explore how groups behave in a virtual environment like \system{} to solve visual analytics tasks, and to understand how users utilise surface-based functionalities to accomplish this.
To do so, we conducted an exploratory study with 10 groups of three participants\deleted{solving data visualisation and analysis tasks together}, split equally between two study parts with five groups each: Part A and Part B. In Part A, we limit participants to 2D visualisations and with only the walls as 2D surfaces. In Part B, we further explore the surface metaphor by including the virtual table, as well as enabling 3D visualisations. Otherwise, study conditions were kept as similar as possible, which we describe in this section.

\subsection{Dataset and Study Design} \label{ssc:study-design}
We used a housing auction result dataset with 8,400 data records collected from the Melbourne, Australia region. This was chosen as it is simple and recognisable to most of our participants, and has a combination of temporal, spatial, categorical, and numerical dimensions. The study focused around two types of data analysis tasks:

\textbf{Directed Tasks (DT).} Groups were given specific questions to answer. This was to both observe how groups achieve a specified goal, and to give some guidance to better familiarise them with \system{} and the dataset. Three of these questions were given, DT1: ``Where is the most expensive house located and why is it so expensive?''; DT2: ``Have house sizes gone up or down over time?''; DT3: ``A family of four is moving to Melbourne with the following requirements... Where would you recommend them to live?'' Groups were expected to give an answer with consensus.

\textbf{Free Exploration Task (FET).} Groups were given 15 minutes to freely explore the same dataset to find any insights they deemed interesting. This was to observe how they worked together towards a more open objective. They then had to present these findings to a virtual embodied audience avatar controlled by keyboard and mouse by the experimenter. The \deleted{audience} avatar was positioned at the direction of the group, and was controlled as passively as possible (i.e. minimal movement, no follow-up questions asked).

\subsection{Procedure}
Each session \replaced{was}{lasted} approximately 90 minutes \added{from start to finish,} and \replaced{were}{was} conducted as follows.

\textbf{Introduction and Training} \textit{(20 min).} Each group was given a brief introduction to the study, followed by a demographics questionnaire. Participants were then trained on how to use \system{} by the experimenter who was present in the virtual environment as a fourth user, allowing them to `follow along' the instructions and become accustomed to having others around them. In Part B, they were also taught how the table and 3D visualisations worked.

\textbf{Directed Tasks} \textit{(30 min).} Groups were given directed task one after the other. The task was written on a specific virtual wall 
and was read out loud. At the beginning, authoring panels were instantiated in intentionally awkward positions to force participants to reposition them manually, but these panels (and all visualisations) were not forcefully reset or removed between tasks. 
Verbal time warnings were given, but much leeway was given for those still actively working on an answer. 

\textbf{Break} \textit{(10 min).} Groups were then given a 10 minute break. During this time, they answered a mid-study questionnaire about their collaborative strategies and behaviours. Participants were not permitted to discuss the questionnaire amongst each other whilst answering it.

\textbf{Free Exploration Task} \textit{(25 min).} After putting on the VR equipment again, groups were given the free exploration task. Instructions and warnings were the same as for the directed tasks.

\textbf{Wrap Up} \textit{(5 min).} Participants were then given a final post-study questionnaire, which was similar to the mid-study questionnaire but with additional questions about immersion in the environment.

\subsection{Participants}
We recruited 30 participants (7 female, 23 male) aged between 18 and 64, with all but one participant having a background in computer science. 74\% were familiar \deleted{(self-reported 3 or higher out of 5) }with data visualisation and 74\% with VR technology \added{(self-reported 3 or higher out of 5)}. The participants were recruited via word of mouth and were a mix of academics and post-graduate students\added{ from our university}. All had normal or corrected-to-normal vision and had no strong susceptibility to motion sickness.
All but two participants reported being familiar with \added{the other members in} their group.

\subsection{Data Collected} \label{ssc:data-collected}
During the study, a fourth PC was used to collect the spatial positions of each participants heads, hands, and of all objects in the virtual room. It was also used to collect the start and end times of all interactions, such as panel button clicks and visualisation interactions. Finally, it was used to video record a top-down view of the virtual room. The questionnaire contained a mix of Likert-scale and short-answer based questions. Likert-scales were on a scale of 1 (disagree) to 5 (agree).

\subsection{Analysis Process}
Three coders initially coded the videos with an open coding methodology looking for collaborative activity and patterns, specifically identifying periods of discussion and tightly-coupled collaboration.
We also processed the log data to analyse participant movement and actions, as well as the placement and properties of objects in the environment.
We then more closely viewed the video footage to extract a deeper qualitative contextual understanding of individual and collaborative behaviours. See supplemental material for further details\footnote{https://sites.google.com/monash.edu/shared-surfaces-and-spaces}.
\section{\replaced{Findings and Observations}{Results}}
\label{sec:results}
We first present general findings about the study and how users utilised \system{}, followed by our main study results structured around a \textit{representative group workflow} described later.
For the rest of this paper, groups are prefixed with a G and participants with a P. G1--G5 and P1--P15 are Part A, G6--G10 and P16--P30 are Part B.

\subsection{General Results} \label{ssc:general-results}
\begin{table}
\centering
\resizebox{\linewidth}{!}{%
\begin{tabular}{cccccccc}
\hline
 &  &  &  & \multicolumn{2}{c}{\textbf{Actions Performed {[}per min{]}}} & \multicolumn{2}{c}{\textbf{Collab. Duration {[}min{]}}} \\ \cline{5-8} 
 & \multirow{-2}{*}{\textbf{Task}} & \multirow{-2}{*}{\textbf{\begin{tabular}[c]{@{}c@{}}Duration\\ {[}min{]}\end{tabular}}} & \multirow{-2}{*}{\textbf{\begin{tabular}[c]{@{}c@{}}Movement\\ {[}m/min{]}\end{tabular}}} & \textbf{Grasping} & \textbf{Ranged} & \textbf{Discussion} & \textbf{Group Work} \\ \hline
\multicolumn{1}{c|}{} & \multicolumn{1}{c|}{\textbf{DT1}} & 6.4 & 6.7 & 4 & 1.8 & 2.0 & 0.9 \\
\multicolumn{1}{c|}{} & \multicolumn{1}{c|}{\textbf{DT2}} & 7.0 & 6.3 & 3.5 & 2.2 & 3.0 & 0.6 \\
\multicolumn{1}{c|}{} & \multicolumn{1}{c|}{\textbf{DT3}} & 13 & 5.8 & 2.6 & 2.8 & 3.5 & 2.6 \\
\multicolumn{1}{c|}{} & \multicolumn{1}{c|}{\textbf{FET}} & 20.2 & 6.2 & 2.8 & 3.1 & 6.5 & 2.4 \\ \cline{2-8} 
\multicolumn{1}{c|}{\multirow{-5}{*}{\textbf{Part A}}} & \multicolumn{1}{c|}{\cellcolor[HTML]{EFEFEF}\textbf{Total}} & \cellcolor[HTML]{EFEFEF}46.7 & \cellcolor[HTML]{EFEFEF}6.2 & \cellcolor[HTML]{EFEFEF}3 & \cellcolor[HTML]{EFEFEF}2.7 & \cellcolor[HTML]{EFEFEF}3.8 & \cellcolor[HTML]{EFEFEF}1.6 \\ \hline
\multicolumn{1}{c|}{} & \multicolumn{1}{c|}{\textbf{DT1}} & 5.9 & 5.1 & 3 & 1 & 1.2 & 0.5 \\
\multicolumn{1}{c|}{} & \multicolumn{1}{c|}{\textbf{DT2}} & 5.4 & 5.2 & 1.7 & 1 & 0.8 & 1.1 \\
\multicolumn{1}{c|}{} & \multicolumn{1}{c|}{\textbf{DT3}} & 12.3 & 4.6 & 1.4 & 1 & 1.1 & 4.3 \\
\multicolumn{1}{c|}{} & \multicolumn{1}{c|}{\textbf{FET}} & 20 & 5.3 & 1.7 & 1.7 & 5.8 & 4 \\ \cline{2-8} 
\multicolumn{1}{c|}{\multirow{-5}{*}{\textbf{Part B}}} & \multicolumn{1}{c|}{\cellcolor[HTML]{EFEFEF}\textbf{Total}} & \cellcolor[HTML]{EFEFEF}43.5 & \cellcolor[HTML]{EFEFEF}5.1 & \cellcolor[HTML]{EFEFEF}1.8 & \cellcolor[HTML]{EFEFEF}1.3 & \cellcolor[HTML]{EFEFEF}2.2 & \cellcolor[HTML]{EFEFEF}2.5 \\ \hline
\end{tabular}%
}
\caption{Averages of task duration, participant movement, actions performed on panels and visualisations, and collaboration time collected from trace data and collaboration video coding.}
\label{tab:average-counts}
\end{table}

Table~\ref{tab:average-counts} shows the averages of task times, lateral head movement, and number of grasping and ranged actions performed on both panels and visualisations.
Groups in Part B took less time than those in Part A, but this was not strictly controlled during the study. Part A performed many more actions per minute than Part B, particularly due to their \replaced{increased}{much higher overall} use of the panel to modify visualisation properties (4000 total actions on panels in Part A, 2037 in Part B).
Participants had mixed preferences of these grasping vs ranged actions: 17 predominantly used grasping actions, 10 mainly used ranged actions, and three used a mix of the two.
To our surprise, only three groups (G2, G3, G4) made \added{substantial} use of the brushing feature\added{ (longer than three minutes continuous brushing), with other groups only using it either briefly or not at all}. In contrast, the details-on-demand tool was frequently used \added{by all}. While intended for inspection of specific points such as outliers, many used it to `scan' over the visualisation, rapidly inspecting many points in quick succession. This was also used to point at specific parts of visualisations.

\begin{figure}[t!]
    \centering
    \includegraphics[trim = 0 5 0 0, clip, width=0.85\linewidth]{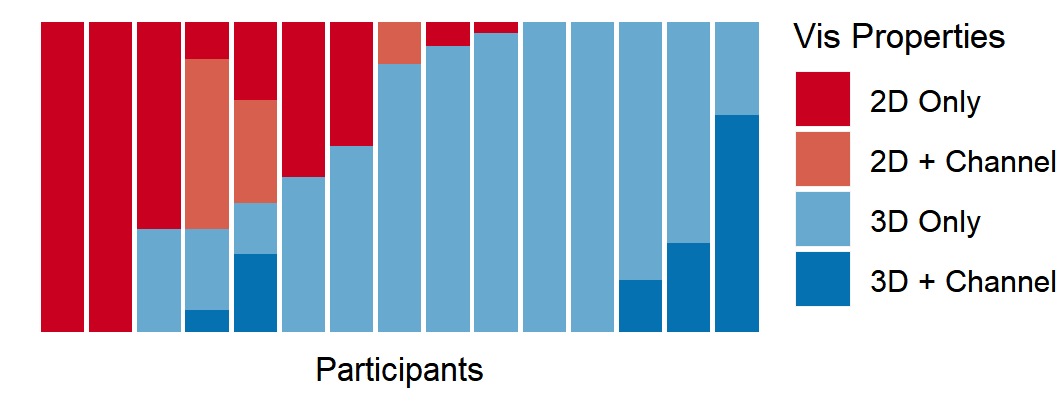}
    \caption{Proportion of 2D versus 3D visualisations teared out per each participant for all tasks in Part B, further split by if an additional channel (size, colour) was present in the visualisation.}
    \label{fig:2d-vs-3d}
\end{figure}

\begin{figure}[b!]
    \centering
    \includegraphics[width=\linewidth]{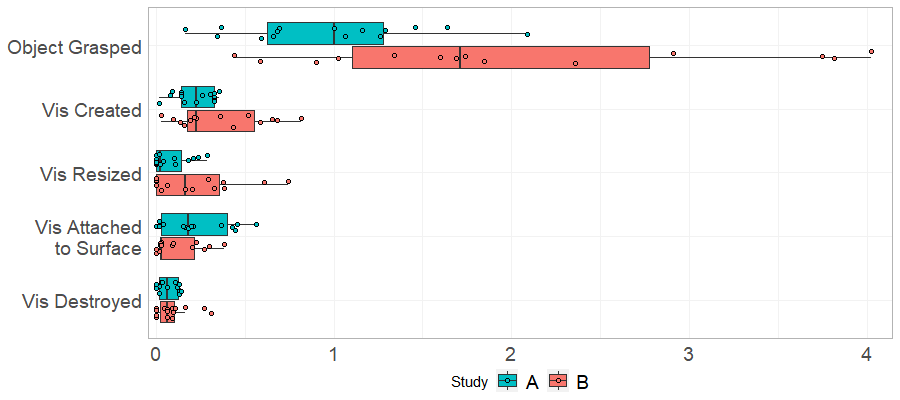}
    \caption{Boxplots of actions per minute for certain interaction types for each participant across all tasks. Vertical lines represent quartiles, points represent individual participants.}
    \label{fig:actions-per-minute}
\end{figure}

In Part B, a considerable number of participants created more 3D than 2D visualisations (Fig.~\ref{fig:2d-vs-3d}). This suggests a willingness to use the $z$-axis to visualise a third dimension instead of other visual channels such as size or colour. However, this was partly \replaced{the result}{a function} of 3D visualisations being easier to manipulate and rotate when torn out than when still on the panel. When comparing the number of creation actions per minute between study parts in Fig.~\ref{fig:actions-per-minute}, those in Part B noticeably created more visualisations than those in Part A, but conversely performed far less actions on the panel to modify visualisation properties (Table~\ref{tab:average-counts}).  

Participants responded positively when asked about their experience with \system{}. They felt immersed in the virtual environment (27 participants reported a 4 or 5), felt comfortable using the VR setup (25 reported a 4 or 5), and that they felt their group-mates were really there with them (25 reported a 4 or 5).
Taking into account setup times and training, participants spent roughly 80 minutes in VR, with the longest continuous duration being 50 minutes. Despite this, the majority of participants reported little to no cybersickness throughout the study (4 participants reported a 3, the rest 1 or 2)\added{, but a few complained that the heat of the VR backpacks had started to become uncomfortable}. \added{No participant commented on any issues regarding usability or latency, however this was not explicitly asked in the questionnaire.}

\newcommand{\alice}{Alice}
\newcommand{\bob}{Bob}
\newcommand{\carol}{Carol}

\subsection{Representative Group Workflow}
\label{ssc:workflow}
We structure our main study \replaced{observations}{results} around a \textit{representative group workflow}. That is, we identify major stages in the groups' workflows and describe each with a narrative (\textit{italic text}) of notable behaviours (with generalised personas \alice{}, \bob{}, and \carol{}). Below each stage, we describe how groups followed or deviated from it, as well as any alternate behaviours we observed.

\vspace{0.5mm}
\textbf{\hypertarget{wf:workspace-organisation}{[Stage 1]} Setting up the workspace.}
\textit{\alice{}, \bob{}, and \carol{}, are working together to find insights in the dataset akin to the free exploration task. They each place their panels along a separate wall without any verbal coordination.
}

\begin{figure*}[h]
    \centering
    \includegraphics[width=0.45\linewidth]{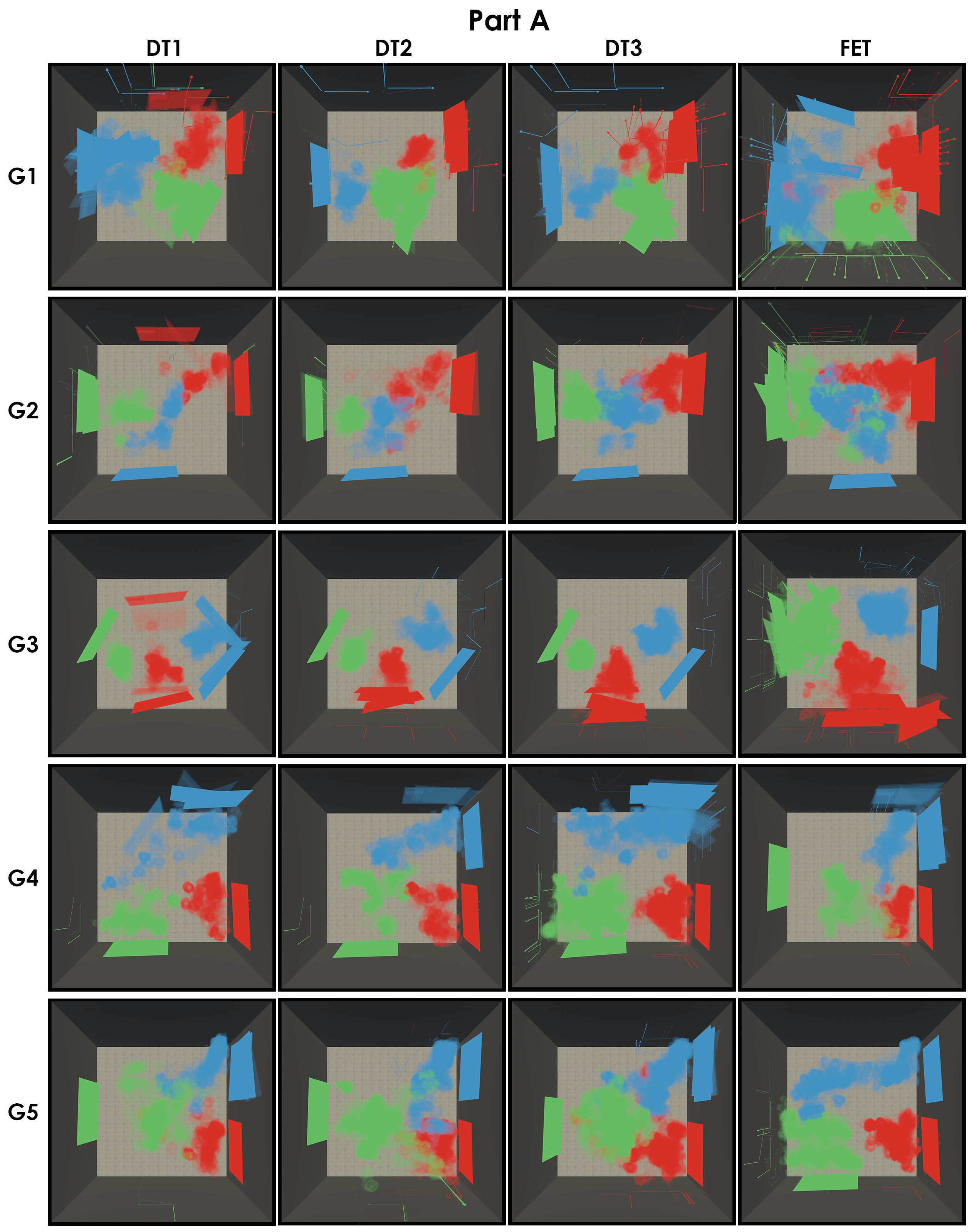}~
    \includegraphics[width=0.45\linewidth]{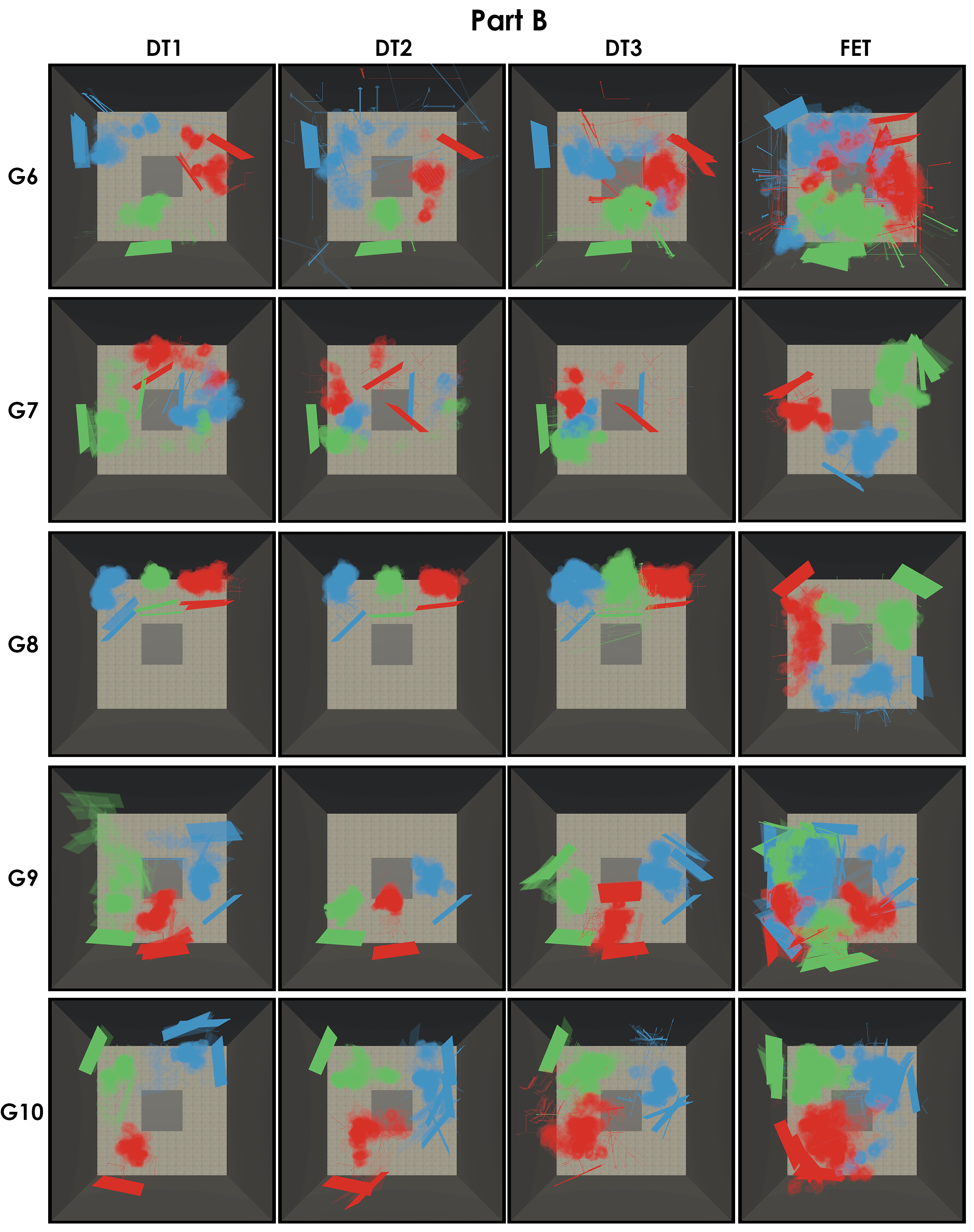}
    \caption{Bird's-eye view heatmaps of each participant and their objects sampled every 5 seconds using a perspective projection. Large dark grey border is the wall, grey centre square in Part B is the table. Brighter areas indicate more time spent there. Red, green, and blue each represent a unique participant in that group, visualisations and panels use the same colour as their owner.}
    \label{fig:heatmaps}
\end{figure*}

Placement of panels and movement of participants throughout all tasks can be seen in Fig.~\ref{fig:heatmaps}.
Many groups did not negotiate where to place panels or how to divide the workspace, but always resulted in roughly equally sized areas for each participant.
Two groups (G8, G9) positioned themselves side-by-side to more easily see each other, with the other groups taking separate parts of the room.
Groups that took separate parts of the room placed their panels against a wall or the table, only moving them for specific reasons such as to move closer to their group members (G9 in Fig.\ref{fig:heatmaps} as an example).
In almost all cases, panel placement was intrinsically linked to where its owner comfortably moved and worked, essentially defining their personal territory. This is noticeable in Fig.~\ref{fig:heatmaps} as participant movement was typically in front of their panel. Outside of tightly-coupled collaboration, no participant entered the territory of another.

\vspace{0.5mm}
\textbf{\hypertarget{wf:collab-strategy}{[Stage 2]} Collaboration strategy.}
\textit{\alice{} suggests formulating a strategy. The group decides to divide the work: \alice{} looks at the most expensive properties in the data, \bob{} at temporal trends, and \carol{} at spatial characteristics.
}

In most cases, groups began exploring the data on their own panels without explicitly discussing strategy.
Even so, they ended up using mixed-focus collaboration styles, seeking assistance and/or sharing findings throughout the task. 
Three groups (G3, G4, G8) progressed toward divide and conquer strategies, particularly during the free exploration task. They would negotiate aspects and dimensions of the data they each wanted to explore before splitting up, occasionally checking their results with each other.
Three other groups (G6, G7, G9) organically shifted to tightly-coupled collaboration for a few tasks in the study, working on a panel or visualisation together for a significant period of time.
Overall, participants would usually break off to work individually on their own panels, but would remain engaged with any ongoing discussion and seamlessly rejoin when needed.

\begin{figure}[!b]
    \centering
    \includegraphics[height=3.45cm]{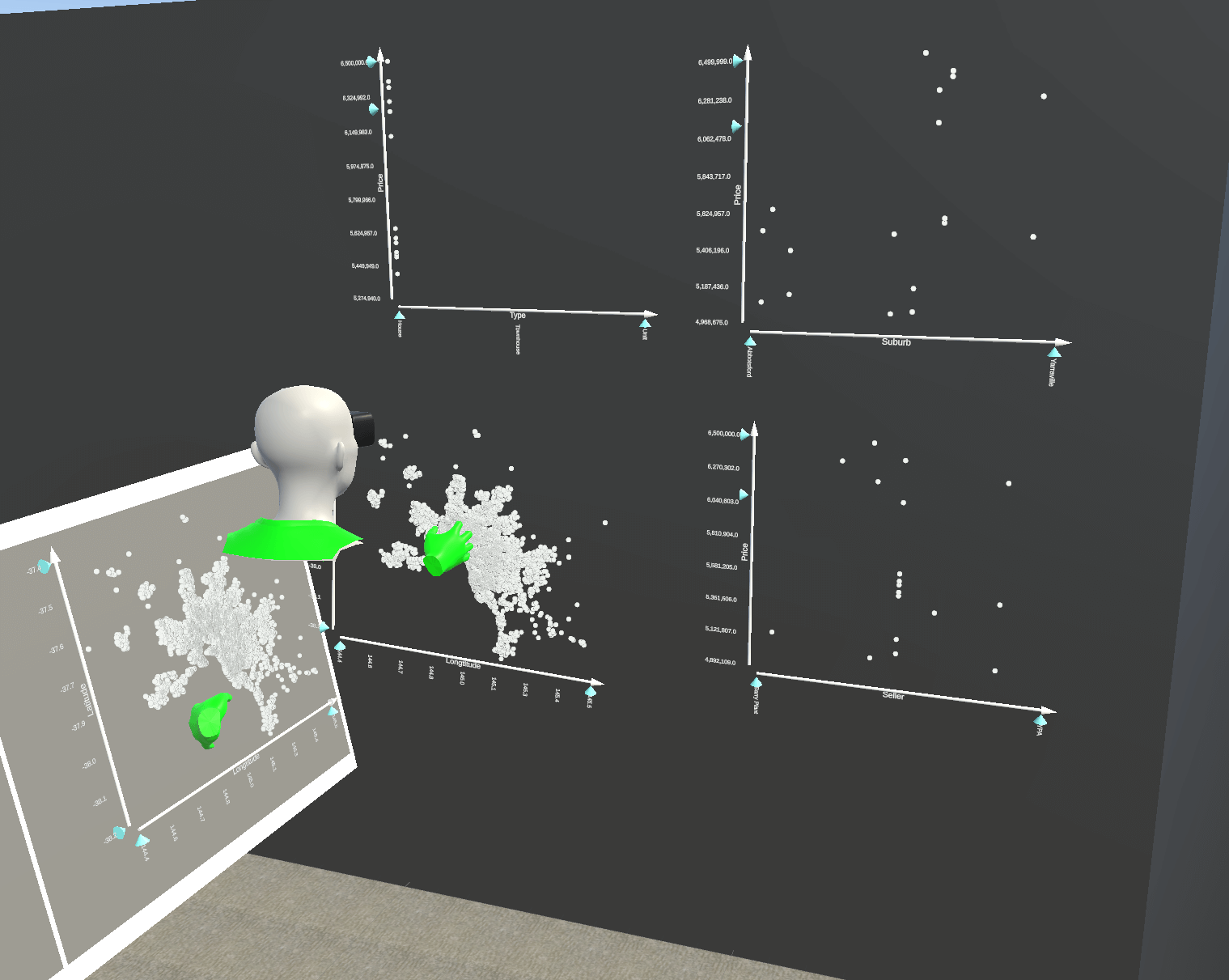}~
    \includegraphics[height=3.45cm]{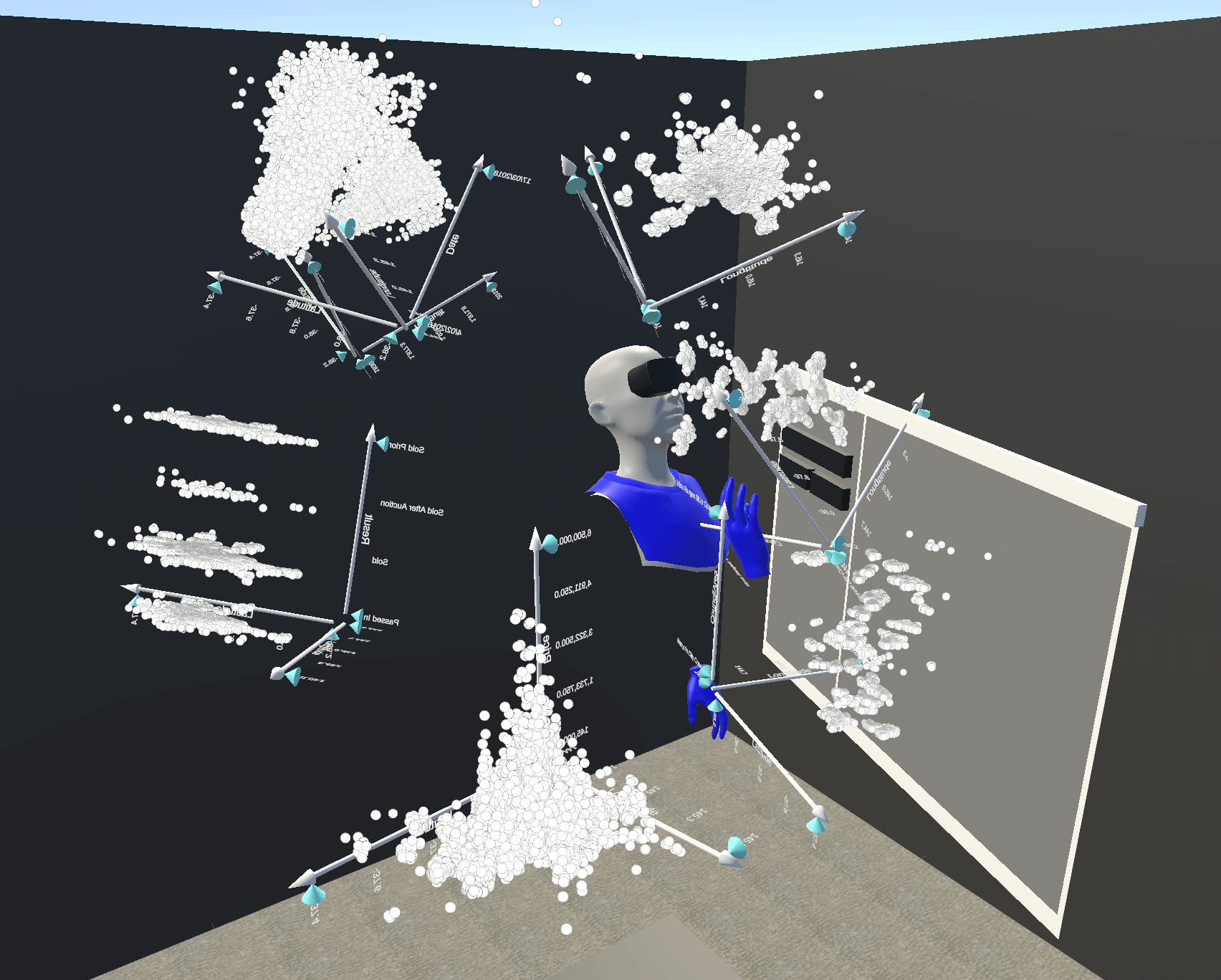}
    \caption{Actual instances of planar \added{used by P6} (left) and egocentric layouts \added{used by P29} (right) used by participants.}
    \label{fig:planar-egocentric}
\end{figure}

\vspace{0.5mm}
\textbf{\hypertarget{wf:individual-work}{[Stage 3]} Working individually.}
\textit{They each work independently in separate areas.
\alice{} isolates outliers along a Price dimension with the axis scalers and cycles through dimensions on the other. She tears out `completed' 2D scatterplots and pins them neatly on the adjacent wall.
\bob{} tears out a 2D scatterplot of Date $\times$ LandSize, widening it to around 2~m along its $x$-axis to visually analyse it in front of his panel.
\carol{} tears out multiple 3D scatterplots of Longitude $\times$ Latitude with various $z$-axis dimensions, rotating each directly in front of her and leaving previous ones in the open space around her.
}

Most participants used two distinct visualisation layouts while working: a planar layout with visualisations in a grid-like fashion, as per \alice{} (8 in Part A, 1 in Part B), or an egocentric layout with visualisations in a spherical-like fashion around them, as per \carol{} (5 in Part A, 12 in Part B). Examples of both are shown in Fig.~\ref{fig:planar-egocentric}. Planar layouts would oftentimes use the wall for 2D visualisations in Part A, exemplified by the higher attach action count in Part A in Fig.~\ref{fig:actions-per-minute}. Note that some participants with egocentric layouts changed to planar layouts for presentation (described in \hyperlink{wf:presentation}{Stage 8}).
We also observed other styles, such as P5 who only placed visualisations in a central `shared' area, and P17 and P18 who would tear out and analyse a single large 2D visualisation, as per \bob{}.
To pan across large visualisations, participants either walked back and forth (P17), or grabbed and moved the visualisation sideways, clutching as many times as necessary (P18, P28) in order to minimise body movement.
It was generally not necessary to move for smaller visualisations. We saw no instances of participants orbiting around 3D visualisations when working independently, instead rotating them on the spot by clutching (hence the visibly higher grasp counts for Part B in Fig.~\ref{fig:actions-per-minute}).

\vspace{0.5mm}
\textbf{\hypertarget{wf:collab-transition}{[Stage 4]} Transitions to tightly-coupled collaboration.}
\textit{\carol{} finds something interesting, and announces to her group that she has something to show them.
}

\begin{figure}
    \centering
    \includegraphics[height=3.68cm]{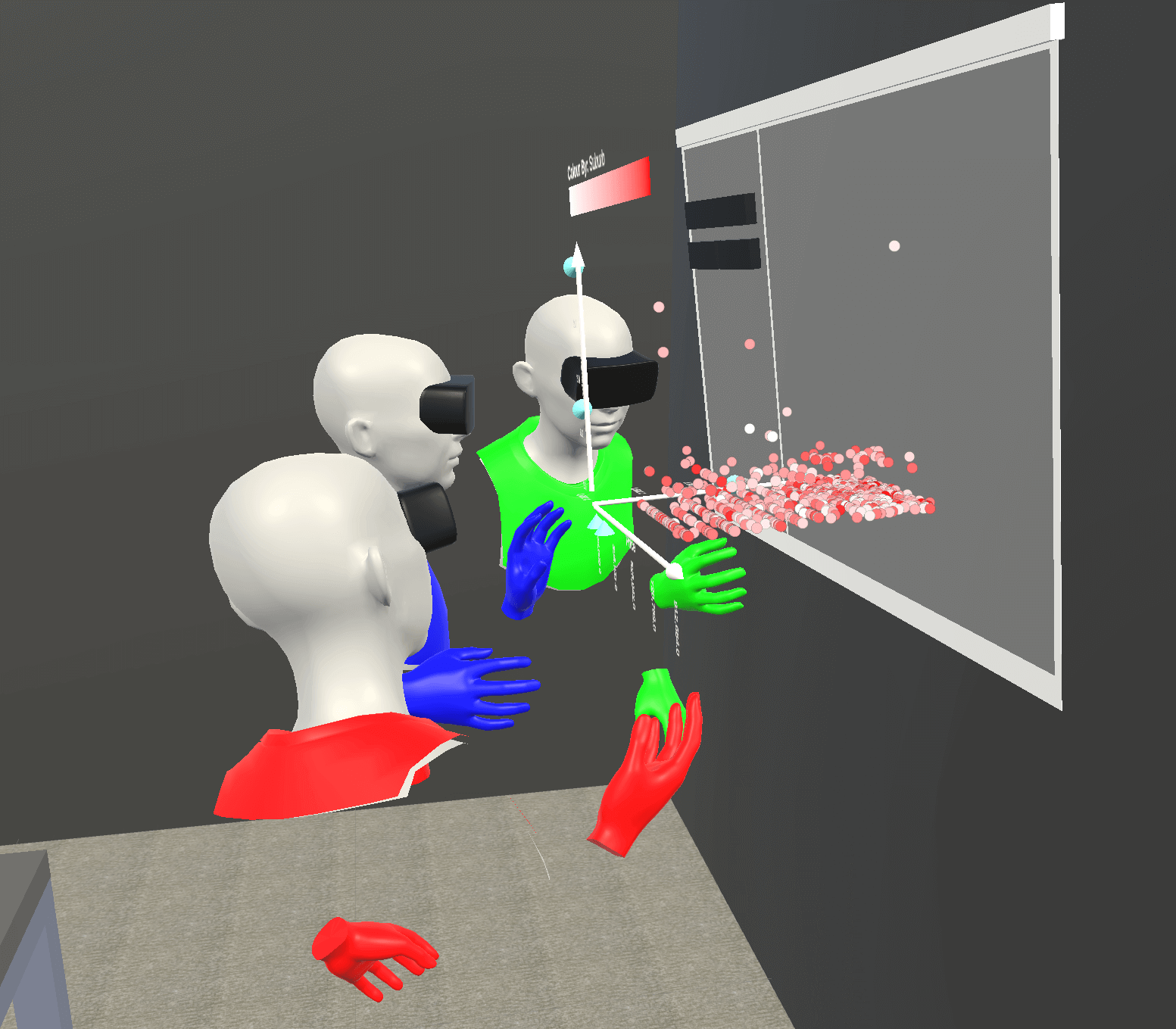}~
    \includegraphics[height=3.68cm]{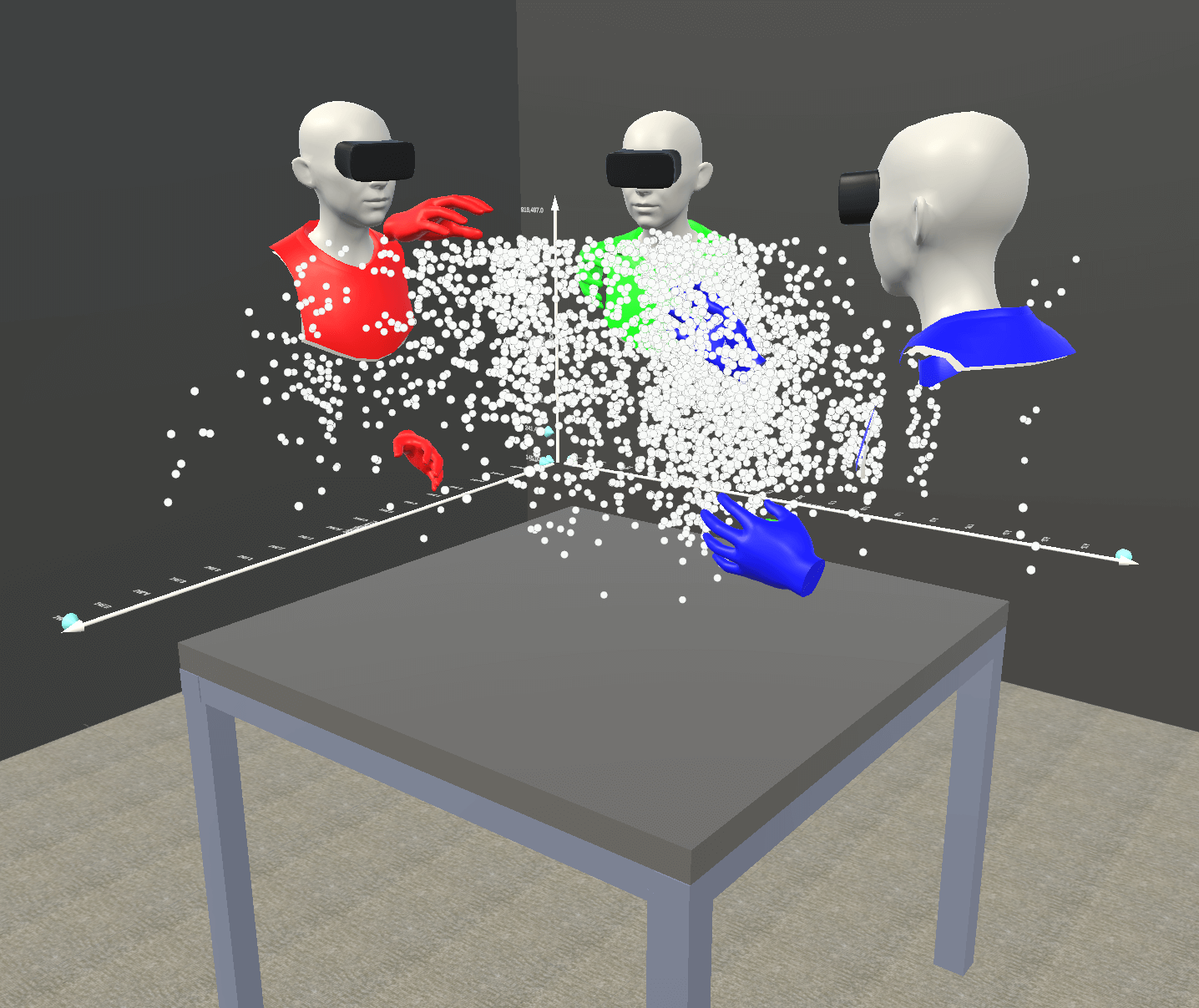}
    \vspace{-3mm}
    \caption{Actual instances of tightly-coupled collaboration: \added{G7} working on the same panel together (left), \added{G6} working on a 3D visualisation on the centre table together (right).}
    \label{fig:tightly-coupled-collab}
\end{figure}

Participants often communicated regardless of physical location to ask for assistance and share findings. In some cases, this lead to tightly-coupled collaboration around a visualisation or a panel (Fig.~\ref{fig:tightly-coupled-collab} left). Sometimes this collaboration was instigated more by action: moving visualisations in front of others (P14, P16), or placing visualisations in a group territory (P1, P4, P16).
This group territory was only present in three groups (G1, G2, G6) and was typically in the centre of the room, acting as a space to analyse visualisations together. This space was never explicitly negotiated or defined. In other groups, participants would either look from their own personal territory (G3, G4, G8, G10), or enter each others' territories only during tightly-coupled collaboration (G5, G7, G10).
The openly shared nature of \system{} allowed for some spontaneous moments of collaboration as well, such as P12 assisting in identifying specific data points from across the room using the pointer, or P25 noticing and offering to help a struggling team member.
When asked if VR helped or hindered collaboration, seven stated it was useful to see each others' work in real-time to improve workspace awareness, and another seven stating that it was easy to share findings with one another. Overall, 21 participants rated a 3 or more for being aware of what their team members were doing at all times.

\vspace{0.5mm}
\textbf{\hypertarget{wf:surface-use}{[Stage 5]} Surface and 3D visualisation usage.}
\textit{\carol{} places a 3D scatterplot of Longitude $\times$ Latitude $\times$ YearBuilt on the table, with the others joining her on opposite sides.
}

\noindent Contrary to our intentions, only G6 placed 3D visualisations on top of the table with the snapping functionality for group work (Fig.~\ref{fig:tightly-coupled-collab}, right). 
Other groups would either place visualisations at head height above the table (G8, G10) or not use the table at all (G7, G9).
Despite this, G6 would sometimes place 3D visualisations floating in space, which was usually a result of the table already being occupied. 
The group would also sometimes struggle with placing the visualisations exactly on top of the table, often needing to make multiple adjustments due to the lack of gravity.
In contrast, participants frequently placed 2D visualisations along the walls, as described earlier for planar layouts, but never placed them on the table.
Tightly-coupled collaboration with 2D visualisation was always done side-by-side. For 3D visualisations, groups positioned themselves around the visualisation (G6, G7, G8, G10) more often than side-by-side (G6, G9). However this positioning was oftentimes a natural consequence of their original starting positions.
Unlike working individually, participants working together with 3D visualisations refrained from rotating them unless contextually appropriate, such as explaining things to others (P23, P28), rotating to help find a specific data point together (P26, P27), or to collectively view it side-on rather than top-down (G6).
There was only one instance where a participant, P20, moved around a 3D visualisation to view it from the same perspective as another.

\vspace{0.5mm}
\textbf{\hypertarget{wf:spatial-awareness}{[Stage 6]} Spatial awareness of virtual objects.}
\textit{They notice a conical shape of recent properties extending outwards over time from the central business district. They walk `through' the virtual table and `into' the visualisation to reach the \replaced{city centre}{central business district}, using the details-on-demand tool to investigate historical properties\deleted{in the city centre}.
}

We did not instruct participants to avoid walking through the table (unlike the VRTable in Kraus \emph{et al.} \cite{Kraus:2020:IIC}), but eight of 15 were observed to deliberately avoid it.
One group of three (G6) did walk through the table in a similar manner to the presented scenario, standing inside of a large 3D visualisation for closer inspection (similar to the VRRoom \cite{Kraus:2020:IIC}) rather than rescaling the visible ranges along visualisation axes.
However, this occurred both on top of and away from the table, using large visualisations ($\approx$1.7~m~$\times$~1.7~m) that were placed in the corners of the room.
The remaining four of 15 participants paid little heed to the table, occasionally walking through it either on accident or when walking from one end of the room to the other.
While most participants did not use the table, some placed their panels against it as if it were a wall (G8 in Fig.~\ref{fig:heatmaps} as an example).
In contrast, participants never walked through walls (as it acted as the room boundary) and all but three avoided moving through panels.
In two cases (G8, G9) this was pushed to the extreme, as they placed panels side-by-side in such a manner where it cordoned off two-thirds of the room, inadvertently restricting their available space.


\begin{figure}[t]
    \centering
    \includegraphics[width=0.95\linewidth]{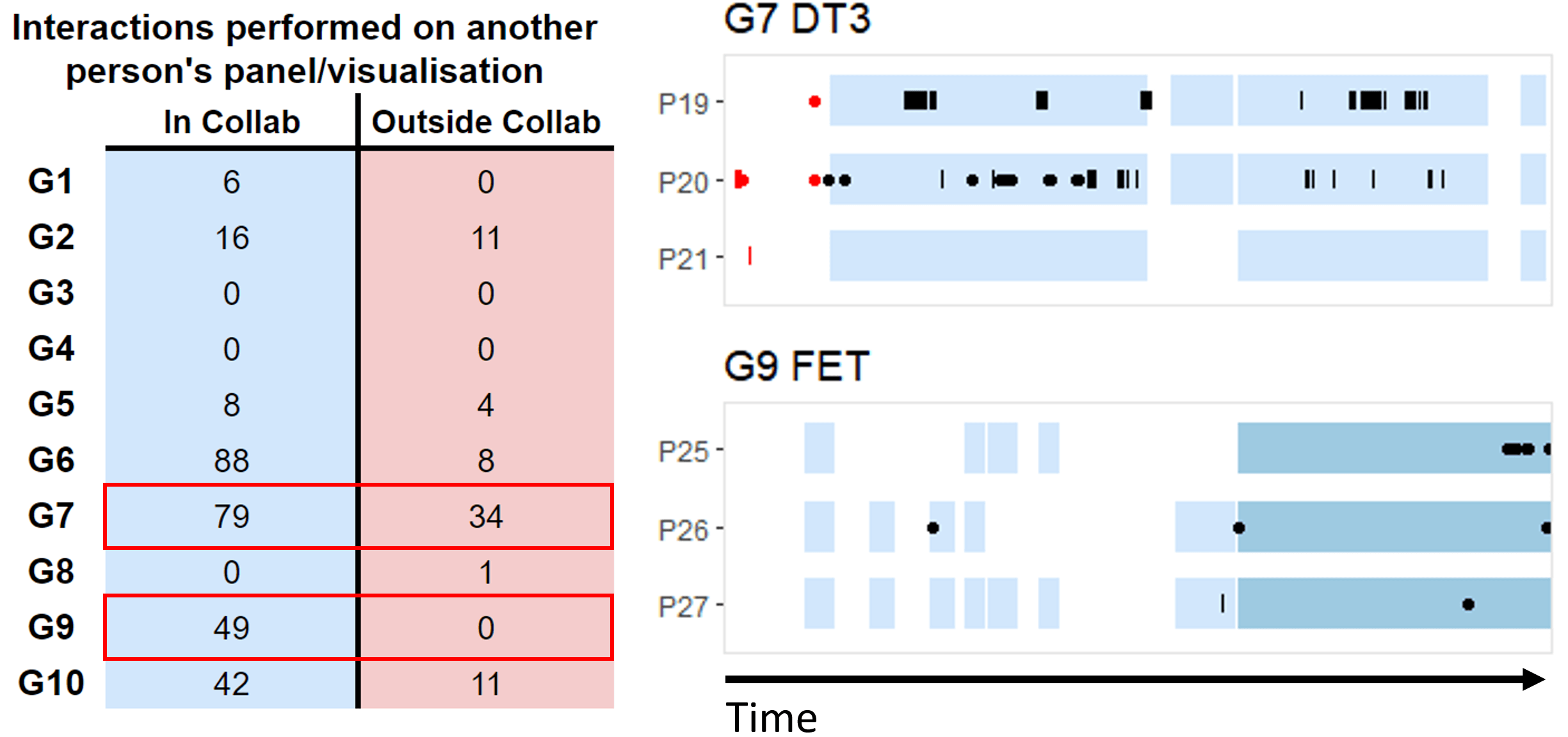}
    \vspace{-3mm}
    \caption{Total counts of interactions performed on objects created by someone else, in and outside of tightly-coupled collaboration (left); and selected examples showing when these occurred in relation to tightly-coupled collaboration and presentation phases (right). Light blue phases are tightly-coupled collaboration, darker blue is presentation in FET.  Black dots and bars are discrete and continuous interactions performed during collaboration, red dots and bars are outside of collaboration. }
    \label{fig:interaction-on-other}
\end{figure}

\vspace{0.5mm}
\textbf{\hypertarget{wf:social-protocols}{[Stage 7]} Behaviours and protocols during collaboration.}
\textit{\alice{} decides to work independently again, pinning 2D visualisations as before. \bob{} draws \carol{}'s attention to his 2D visualisation, stating there is no trend between Date and LandSize. \carol{} suggests that he should change Date to YearBuilt. \bob{} throws the visualisation onto an empty wall, and quickly creates a new YearBuilt $\times$ LandSize scatterplot. Observing the slight positive trend, \carol{} brushes the largest properties to highlight them on the 3D scatterplot. While this selection also shows on \alice{}'s visualisations, she makes no fuss about it.
}

Those in Part B were much more willing to interact with panels and visualisations belonging to others than in Part A. This was more likely to occur during tightly-coupled collaboration (Fig.~\ref{fig:tightly-coupled-collab}), such as when working on the same panel together (G7, G9) or to adjust visualisation size of a central 3D visualisation (G6). For occurrences outside of tightly-coupled collaboration, it was either during presentation in FET, or at the start and end of the task when groups manually reset the room of all visualisations (Fig.~\ref{fig:interaction-on-other}).
Some participants, particularly those in Part A, would instead suggest or direct the owner to perform actions for them, for example P6: ``Along the longitude latitude graph, colour by result on the map...'' P4: ``...let me just do it for you...'',
but three participants were willing to physically reach over someone else to modify a panel/visualisation.
When not manipulating visualisation properties, participants made constant use of the pointer (accessed through details on demand) during their discussion to highlight specific parts of visualisations.
Ultimately, objects would belong to the participant who created it, with others avoiding interacting with said objects except during tightly-coupled collaboration---and in certain circumstances such as those stated in \hyperlink{wf:surface-use}{Stage 5}. Only in G5 was this `ownership' transferred: one instance was implicit with P15 effectively taking control of P13's visualisation during group work, the other explicit with P15 offering P13 to use his panel.

Excluding talking over each other, all interruptions were accidents resulting from the shared nature of the system. For example, P6 being surprised when P5 starts to brush unannounced (as per \alice{}); P9 accidentally erasing P8's selection when performing opposing actions simultaneously; and P28 creating a very wide 3D visualisations which protrudes into P30's workspace. However, in all cases participants made little to no fuss.
In fact, we noticed some would change their behaviour to minimise/avoid conflict, such as P9 only using private brushing after accidentally erasing someone else's selection in a previous task.


\vspace{0.5mm}
\textbf{\hypertarget{wf:presentation}{[Stage 8]} Presentation.}
\textit{For presentation, they decide to organise their visualisations on a shared wall, clustering their visualisations by author. \carol{} however places some of her 3D visualisations close to the wall, with the main one left on the table. With the remotely controlled audience present, they take turns presenting their findings.
\alice{} grabs a 2D visualisation off the wall and brings it closer to the audience, describing it while maintaining eye-contact and using gestures for each visualisation.
\bob{} instead stands a distance from his large 2D visualisations, using a laser pointer to point during his presentation and periodically turning to the audience.
\carol{} moves to each of her 3D visualisations, using the marker to highlight trends and clusters of interest. However, she pays no attention to the audience, and neglects to account for accurate viewing angles.
}

Groups presented an average of 7.5 findings (M = 8, SD = 2.46) to the audience during FET.
Two groups (G1, G9) took time to lay out their findings neatly along a shared set of walls. Groups that did not do this generally already had visualisations grouped in each workspace---particularly for those with planar layouts. Of the 17 who used egocentric layouts, six switched to planar layouts for presentation, nine only had a few visualisations and mainly used their panels, while the remaining two left their (mostly 3D) visualisations in the centre of the room.
G6 was a major exception with many visualisations scattered around the area---a result of their analysis involving many large 3D visualisations.
As a result, P16 lead his group's presentation for the majority of their findings. All other groups took turns presenting (excluding P1, P2, P5 who did not present). While no presentation had any overarching story, some focused on a specific theme as a result of their division of work, such as P9 \replaced{focusing}{foucusing} on most expensive properties and P22 with comparison between houses, units, and apartments.

\begin{figure}
    \centering
    \includegraphics[height=2.9cm]{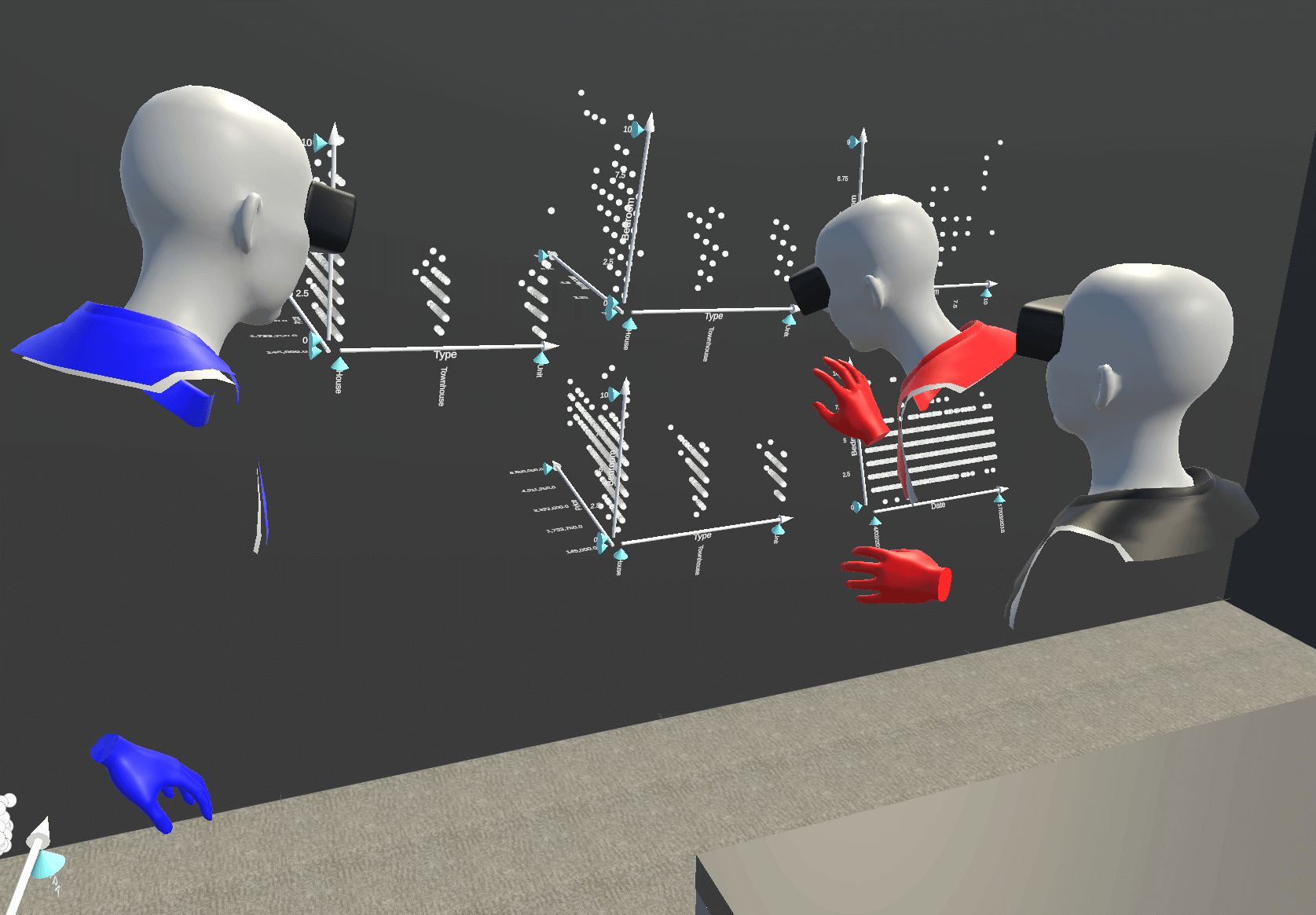}~
    \includegraphics[height=2.9cm]{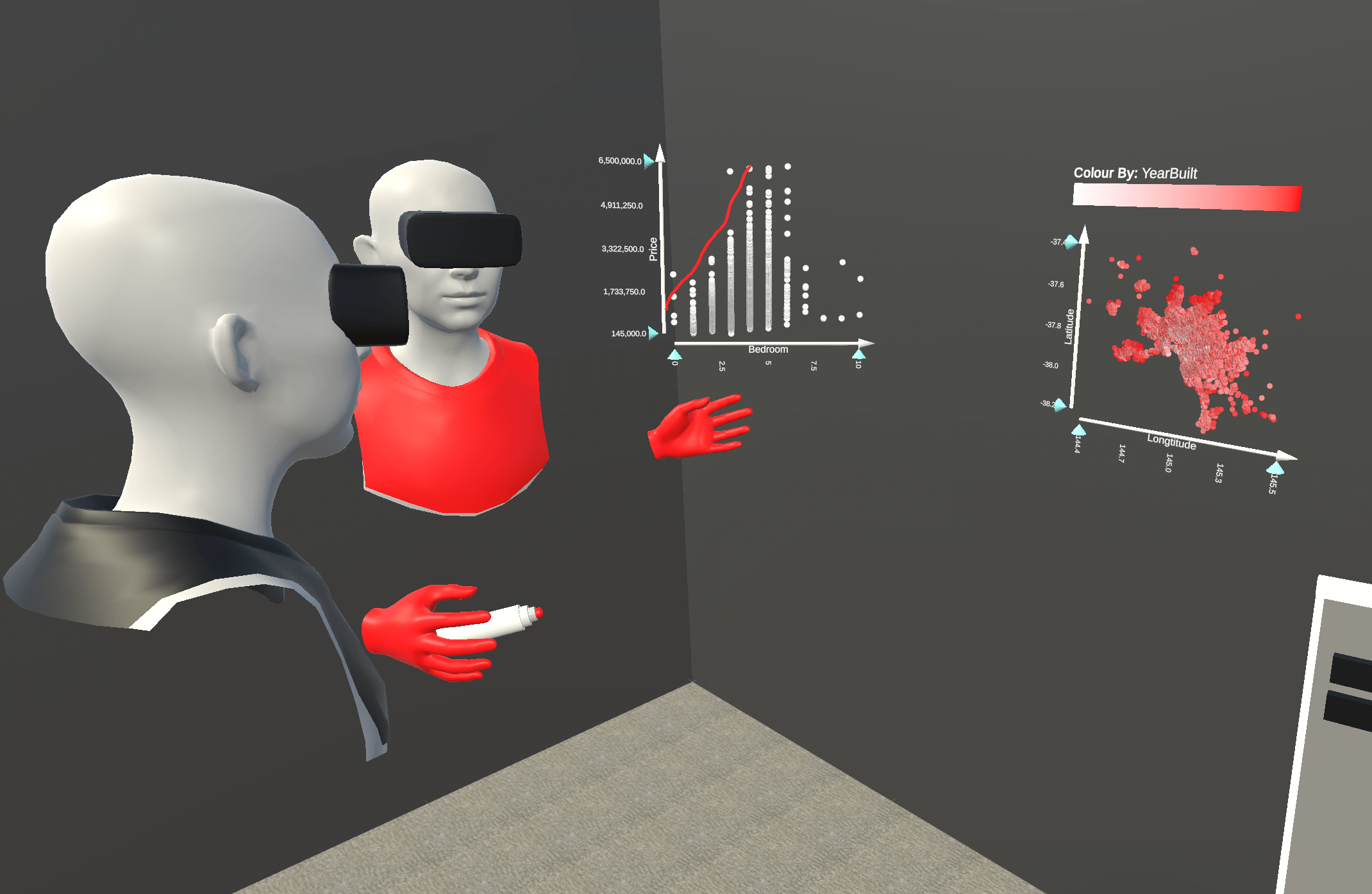}
    \vspace{-3mm}
    \caption{
    Actual examples of presentation styles in FET: P22 (in red) explaining insights on 3D visualisations without accounting for the audience's perspective, with his back turned towards the audience (in black) (left); P25 using the marker to highlight a trend on a visualisation while exhibiting proper body language to the audience (right).}
    \label{fig:presentation-styles}
\end{figure}

While presenting, 16 used gestural hand movements (as per \alice{}), six used the laser pointer (as per \bob{}), four used a mix of the two, and one relied on the marker (as per \carol{}). While the majority in Part B used gestures (3 in Part A, 13 in Part B), this may be because the pointer is disabled when pointing at 3D visualisations.
We also observed that 15 participants had poor awareness of the audience, paying no attention to it while presenting (Fig.~\ref{fig:presentation-styles} left). In contrast, 12 made sure to make eye-contact and engage the audience akin to real life (Fig.~\ref{fig:presentation-styles} right). Participants did not appear to be influenced by the presentation styles of others, such as P8 presenting in a very charismatic tone, and P9 following with a very static/rigid tone.
While three participants grabbed visualisations off walls to bring them closer to the audience,
none made sure to accommodate for the audience's perspective for 3D visualisations, despite it being controlled as passively as possible.
In fact, two participants (P22, P24) reported findings using a 3D visualisation as `two 2D scatterplots in one' akin to the rolling dice metaphor \cite{Elmqvist:2008:RDM}, although neither rotated them to properly face the audience.

\section{Discussion} \label{sec:discussion}
Our study revealed how groups behave in the \system{} collaborative immersive analytics environment. We also observed users solve visual analytics tasks in novel ways made possible by the immersive environment. 
Based on our results, we highlight and discuss important takeaways which can help guide future work in this area.

\subsection{System and Space Usage}
\bpstart{Collaborative visual analytics is possible in immersive environments}
Overall, our participants were capable of performing visual analytics (Sec.~\ref{ssc:workflow}).
    They explored a data set, authored visualisations, discovered insights, organised visualisations in the space around them, and presented their findings to others---
    doing so both independently and collaboratively through mixed-focus collaboration depending on the given context \cite{Gutwin:1998:DID} (\hyperlink{wf:collab-strategy}{Stage 2}).
    They were engaged with the tasks in VR for 45 minutes on average (plus 15 minutes training),
    and made use of most of the functionalities provided.
    However, \system{} is a research prototype, and to observe more complex workflows
    functions such as textual queries and filters, as well as more advanced collaborative support such as privacy management or token-based protocols would be needed. We see this as a logical next step in collaborative immersive analytics to understand how groups utilise such features. Furthermore, we would like to observe how users use \system{} in AR as compared to VR, particularly when integrating with real-world surfaces and proper face-to-face collaboration.

\bpstart{\replaced{Users willingly made use of 3D visualisations, but these require UI designs that support their use}{3D visualisations were frequently used, but required more iterations than 2D in our system}}
    Although we did not emphasise the use of 3D visualisations during training, many of our participants chose to use and experiment with them (Sec.~\ref{ssc:general-results}).
    Some groups naturally came up with interesting ways of viewing these 3D visualisations, such as standing inside of \deleted{an almost} room-scale 3D visualisation\added{s} \replaced{for}{to achieve} an egocentric view (within the data) \replaced{up close}{close up} to specific points \cite{Donalek:2014:ICD, Ivanov:2019:AWA, Kraus:2020:IIC},
    or looking from an axis-aligned viewpoint to view two dimensions\replaced{, and then}{ and} rotating \deleted{them} 90 degrees to view the other two dimensions (similar to a manual \textit{Scatterdice} \cite{Elmqvist:2008:RDM}) (\hyperlink{wf:surface-use}{Stage 5}).
    Participants created many \replaced{more 3D than 2D}{3D} visualisations (Fig.~\ref{fig:2d-vs-3d})\replaced{ as a result}{, mainly because} of the authoring interface we provided. When exploring the data through iterative visual design (e.g.\ changing axes dimensions or aesthetics), participants had to tear out 3D visualisations from the panel to inspect them from different angles. This \deleted{process} resulted in the creation of redundant visualisations which may have impeded their flow, as \replaced{compared to 2D visualisations which could be iterated through}{those in Part A were able to modify and explore a wider range of visualisation variables} at a much faster rate (Sec.~\ref{ssc:general-results}).
    \replaced{While some of these behaviours are a consequence of the panel-based interface, they demonstrate the need for consistent UI designs that equally support both 2D and 3D visualisations---particularly when using 2D WIMP metaphors.}{This observed behaviour leads us to consider alternative ways to author 3D visualisations which would require less iteration, such as more closely embodying visualisation design into the artefact itself \cite{Cordeil:2017:IIA}.}
    \replaced{That said}{Ultimately}, while there was evident interest, engagement, and experimentation with 3D visualisation by our participants, it is difficult to say from our study design whether the use of 3D visualisations had a positive impact on analytic performance. Further study would be required to see if this translates to improved task performance and if the interest persists as people become more experienced with the system.

\bpstart{The presence of 3D visualisations fundamentally changes view management and organisation}
    With only 2D visualisations available in Part A, half of participants placed them neatly on the wall while working (\hyperlink{wf:individual-work}{Stage 3}). This careful placement required a little more physical effort than the egocentric layouts used by other participants and in related work \cite{Cordeil:2017:IIA, Batch:2020:TNS}, but made their work easily visible and presentable to others at all times.
    In contrast, when 3D visualisations were available in Part B, almost all participants used egocentric layouts. This may be due to convenience, the wall affordance not being as suitable for 3D visualisations, or due to perspective issues with 3D making it impractical to accommodate others' viewpoints at all times.
    The introduction of 3D also affected how 2D visualisations were organised, as the same egocentric layout was used for 2D visualisations as well.
    Overall, these \added{are novel} observations \added{that} suggest \deleted{that} view management and organisation---not just analytic performance---can vary greatly based on the types of visualisations present, as the inclusion of 3D visualisations had a profound impact on how participants worked. 
    Future collaborative immersive analytics systems may need to provide different placement tools to accommodate 2D versus 3D visualisations.
\bpstart{Users are influenced by environment configuration only if there is a tangible benefit}
    In contrast to previous work which would situate the task in an `endless void' \cite{Batch:2020:TNS} or tightly couple interaction to a visible surface \cite{Butscher:2018:CTO, Cavallo:2019:IIH, Filho:2019:CIA}, the use of surfaces in \system{} is a completely optional part of users' workflows.
    As previously stated, participants took advantage of the walls in the environment to neatly organise their 2D visualisations throughout each task, whereas previous studies observed users doing so only for presentation \cite{Batch:2020:TNS}.
    In contrast, only a single group made careful use of the table (\hyperlink{wf:surface-use}{Stage 5}), despite it being a common metaphor in previous work \cite{Butscher:2018:CTO, Cavallo:2019:DRH, Kraus:2020:IIC}. 
    This was not for a lack of awareness, as most participants did acknowledge the presence of the table (\hyperlink{wf:spatial-awareness}{Stage 6}).
    Simply put, our participants generally saw no benefit in using the table in its current state, as it adds an unnecessary constraint (e.g.\  height in space) in an environment where visualisations can be placed anywhere\deleted{in space}. This raises \replaced{new questions}{some consideration} as to the value of the tabletop metaphor in collaborative immersive analytics environments, as existing systems \added{using physical tabletops} have \deleted{typically} enforced \replaced{their}{its} use \cite{Butscher:2018:CTO, Filho:2019:CIA}. It may be that the role of tabletops is first and foremost to provide tangible and more comfortable interaction\deleted{as per prior work}---something which the virtual table \deleted{inherently} does not provide. However, introducing \replaced{unique}{additional} functionality may \replaced{promote its use}{make it more useful}, such as \deleted{offering} an alternate authoring interface better suited for 3D visualisations.

\subsection{Collaboration}
\bpstart{Equally distributed interaction resources promotes parallel work and mixed-focus collaboration}
    We chose to provide each person their own authoring panel during the study.
    As a result, we observed that almost all groups utilised mixed-focus collaboration styles \cite{Gutwin:1998:DID}, either working individually and sharing findings throughout the task, or alternating between tightly-coupled and parallel work to try ideas on their own. Only in three tasks (G7 in DT2 and DT3, G9 in FET) did participants work closely with one another for the entire duration (\hyperlink{wf:collab-strategy}{Stage 2}).
    This is consistent with previous findings that users work more in parallel when given the ability to do so  \cite{Birnholtz:2007:ESI}. 
    While future work may evaluate the effects of \replaced{having}{a} single versus multiple authoring interfaces in collaborative immersive environments, a factor to consider is the virtuality of VR/AR. Unlike setups with large physical displays, there is no additional cost for virtual displays and interfaces. While a few groups did shift to a tightly-coupled single display context, the ability for mixed-focus collaboration did appear to be invaluable\deleted{ to many others}\replaced{, allowing groups to work in any manner that best suits them.}{. However, this flexibility is promising for collaboration as groups can easily work in any manner that suits them.}
    

\bpstart{Territories in free-roaming immersive environments are spaces where users can work individually, defined by movement patterns and placement of artefacts}
    Movement of participants and placement of visualisation artefacts appeared to be tightly linked with territorial behaviour. These territories acted as an area where one could comfortably work individually without physical intrusion or disturbance (Fig.~\ref{fig:heatmaps}). Territories were not negotiated, but were typically defined by the initial placement of panels, and participants never entered a territory of another unless for tightly-coupled work (\hyperlink{wf:workspace-organisation}{Stage 1}).
    This behaviour is similar to that found on tabletops \cite{Tang:1991:FOS, Kruger:2004:ROT, Tang:2006:CCO} and wall-sized displays \cite{Jakobsen:2014:UCP}, as these territories were oftentimes respected.

\bpstart{Participants did not interact with objects that did not belong to them}
    We observed that participants kept track of and respected ownership of visualisations and panels (\hyperlink{wf:social-protocols}{Stage 7}), refraining from interacting with artefacts which belonged to others. 
    Instances where participants interacted with someone else's visualisations mainly occurred during tightly-coupled collaboration (Fig.~\ref{fig:interaction-on-other}). Some groups would direct the owner to perform some action for them, while others would simply reach over and do it themselves (\hyperlink{wf:social-protocols}{Stage 7}). The latter was always when the purpose was known to all \cite{Tang:2006:CCO, Jakobsen:2014:UCP}, such as making a visualisation easier for everyone to see, or when the group was already working closely together for an extended period of time. \replaced{This behaviour may also have been influenced by participants being extra cautious of collisions in co-located VR however.}{    This difference in behaviour may also be due to users being careful not to collide with one another due to the use of VR and virtual avatars however.}
    \replaced{While personal territories used similarly to that of previous work \cite{Scott:2004:TCT}}{While personal territories of participants were similar in almost all cases}, group territories were seldom used, acting more as an area in the centre of the room to congregate and discuss rather than to share resources \added{due to the virtuality of the environment}.
    \replaced{Overall, some behaviours are consistent with physical environments, and others unique to co-located immersive environments.}{Many of the observed behaviours exist in a physically co-located immersive analytics environment.} Future work may seek to understand how users collaborate when physically remote, but virtually co-located.
    


\bpstart{Participants favoured workspace awareness over privacy}
    \system{} contains numerous design choices \replaced{to improve}{for} workspace awareness \cite{Gutwin:2002:DFW}, such as the completely shared nature of the environment\deleted{allowing participants to easily look at each others' work from afar}.
    \replaced{Contrary to our expectations}{Although we thought this may be distracting}, many participants commented that it helped improve their awareness of others with only a few reporting it was distracting,
    which allowed for mixed-focus collaboration based on each others' actions (\hyperlink{wf:collab-transition}{Stage 4}) to frequently transition to tightly-coupled collaboration \cite{Gutwin:2002:DFW, Isenberg:2012:CCV, Jakobsen:2014:UCP}.
    \deleted{Awareness was also supported by laser pointers (particularly details-on-demand), as participants could point from across the room and/or identify points of interest up close.}
    \added{The inclusion panel-based authoring tools in \system{} showcased unique and interesting behaviours as well.}
    While participants could turn their panels away from each other to achieve privacy, no \replaced{one}{participants} intentionally did this, with some groups instead trying to optimise their visibility of each other (\hyperlink{wf:workspace-organisation}{Stage 1})\replaced{ or simply}{. Conversely, a few groups} gave up privacy altogether by working on the same panel (\hyperlink{wf:collab-strategy}{Stage 2}).
    It is apparent that this flexibility allows groups to balance the level of awareness and privacy they have, and that immersive environments have different considerations for awareness than conventional virtual workspaces \cite{Gutwin:1998:DID}---namely that users can see much more of the workspace.
    With this in mind, future work may explore alternate forms for improving awareness or privacy management (e.g. \cite{Butz:1998:VMP}) for collaborative immersive analytics.

\bpstart{Ensuring proper perspective is challenging during presentation of 3D visualisations}
While 2D visualisations can accurately be viewed from a wide range of angles, 3D visualisations are reliant on perspective in shared contexts---particularly during presentation.
    However, despite an audience embodied by avatars, presenters did not accommodate well for their audiences' viewpoint with 3D visualisations (\hyperlink{wf:presentation}{Stage 8}).
    This was not for a lack of willingness to engage with the audience, as we observed them actively use gestures and eye-contact.
    \replaced{To the best of our knowledge, only a study by Batch \emph{et al.} had involved presentation to a third-party in an immersive analytics environment \cite{Batch:2020:TNS}. While participants did account for viewpoint, the audience had the same view as the participant themselves, rather than using a separate embodied audience.}{While participants accounted for viewpoint in a similar study involving presentation \cite{Batch:2020:TNS}, they did not utilise a separate embodied audience. Instead, the audience had the same view as the participant themselves, avoiding many of the aforementioned issues.}
    Future work may explore how to facilitate this presentation, whether through an embodied audience with agency, or simply by mirroring the presenter's point of view.
\section{Conclusion \added{and Future Directions}}
\replaced{
Our \system{} system and study is the first to test the effect of surfaces and spaces on visualisation tasks performed by groups collaborating in a room-sized immersive environment. We observed novel behaviours that are unique to collaborative immersive analytics:

\vspace{0.25mm}
\finding Participants reported that the shared VR environment was useful in maintaining workspace awareness and to share findings with each other.  

\finding The authoring panel metaphor allowed them to easily create 2D visualisations and organise them neatly along the virtual walls not only for presentation, but in some cases for exploration and analysis,

\finding 3D visualisations were frequently created, but were freely suspended in convenient locations in space rather than on top of the virtual table.

\vspace{0.25mm}
These surfaces, while optional for users to use, were inherently predefined. A future direction could be to explore emergent workspace configurations when users are able to arrange their own surfaces.
Our study also confirms previous results from collaborative work studies in non-immersive environments:
\vspace{0.25mm}

\finding Groups organically divided the shared virtual environment into equally sized territories which were used for individual work.

\finding They transitioned to tightly-coupled collaboration either by observing each others' work or through discussion regardless of physical distance.

\finding The use of avatars and pointers facilitated collaboration, with deixis allowing participants to work while up close or when far apart.

\vspace{0.25mm}
However, a potential factor in how groups worked was the use of co-located VR, as participants may have been apprehensive to perform certain actions due to the risk of collision. \system{} already supports and has been recently tested with distributed collaboration, thus a natural progression to our work is to explore how users collaborate in distributed immersive analytics environments. Particularly, many social protocols may not exist when collaborators are physically remote, while offering new possibilities such as shared perspective \cite{Cordeil:2017:ICA}.
}
{
Our \system{} system and study is the first to test the effect of surfaces and spaces on visualisation tasks performed by groups collaborating in a room-sized immersive environment.
Participants reported that the shared VR environment was useful in maintaining workspace awareness and to share findings with each other.
The authoring panel metaphor worked well for this, as they could easily create 2D visualisations and place them along the virtual walls during exploration, analysis and presentation. 
3D visualisations were also frequently created, but were freely suspended in convenient locations in space, rather than placing them on the table.
These surfaces, while optional for users to use, were inherently predefined. A future direction could be to explore emergent workspace configurations when users are able to arrange their own surfaces.
Groups easily divided the shared virtual environment into territories for individual work, and were capable of transitioning to tightly-coupled collaboration either by observing each others' work or simply through discussion. The use of avatars and pointers facilitated this, with this deixis allowing participants to work while up close or when far apart.
However, a potential factor in how groups worked was the use of co-located VR, as participants may have been apprehensive to perform certain actions due to the risk of collision. \system{} already supports \deleted{and has been recently tested with}\replaced{remote users}{distributed collaboration}, thus a natural progression to our work is to explore how users collaborate in distributed immersive analytics environments. Particularly, many social protocols may not exist when collaborators are physically remote, while offering new possibilities such as shared perspective \cite{Cordeil:2017:ICA}. 
}

\acknowledgments{
\added{Thank you to our study participants and anonymous reviewers for their feedback.
This research was supported under the Australian Research Council’s Discovery Projects funding scheme (project number
DP180100755) and by an Australian Government Research Training Program (RTP) Scholarship.}
}

\bibliographystyle{abbrv-doi}

\bibliography{bibliography}
\end{document}